\documentclass[aps,prl,floatfix,twocolumn,showpacs,nofitinbib,hyperref=pdftex,citeautoscript,superscriptaddress]{revtex4-1}
\usepackage{epsfig}
\usepackage{amsmath}
\usepackage{amssymb}
\usepackage{amsfonts}
\usepackage{ dsfont }
\usepackage{ comment,color }
\usepackage{lipsum}
\usepackage{hyperref,mathtools}
\usepackage{makecell}
\usepackage{array}
\usepackage{placeins}
\usepackage{pifont}
\usepackage{siunitx}
\hypersetup{colorlinks=true,linkcolor=blue,anchorcolor=blue,citecolor=blue,filecolor=blue,urlcolor=blue,bookmarksnumbered=true,pdfview=FitB}

\newcommand{\beqa}{\begin{eqnarray}}
\newcommand{\eeqa}{\end{eqnarray}}

\begin{document}

\hsize\textwidth\columnwidth\hsize\csname@twocolumnfalse\endcsname

\title{Parametric exploration of zero-energy modes in three-terminal InSb-Al nanowire devices}
\author{Ji-Yin Wang}
\email{wangjiyinshu@gmail.com}
\address{QuTech and Kavli Institute of Nanoscience, Delft University of Technology, 2600 GA Delft, The Netherlands}

\author{Nick van Loo}
\author{Grzegorz P. Mazur}
\author{Vukan Levajac}
\author{Filip K. Malinowski}
\author{Mathilde Lemang}
\author{Francesco Borsoi}

\address{QuTech and Kavli Institute of Nanoscience, Delft University of Technology, 2600 GA Delft, The Netherlands}

\author{Ghada Badawy}
\author{Sasa Gazibegovic}
\address{Department of Applied Physics, Eindhoven University of Technology, 5600 MB Eindhoven, The Netherlands}

\author{Erik P.A.M. Bakkers}
\address{Department of Applied Physics, Eindhoven University of Technology, 5600 MB Eindhoven, The Netherlands}

\author{Marina Quintero-Pérez}
\address{QuTech and Kavli Institute of Nanoscience, Delft University of Technology, 2600 GA Delft, The Netherlands}
\address{Netherlands Organisation for Applied Scientific Research (TNO), Delft 2600 AD, Netherlands}

\author{Sebastian Heedt}
\address{QuTech and Kavli Institute of Nanoscience, Delft University of Technology, 2600 GA Delft, The Netherlands}

\author{Leo P. Kouwenhoven}
\address{QuTech and Kavli Institute of Nanoscience, Delft University of Technology, 2600 GA Delft, The Netherlands}

\date{\today}

\vskip1.5truecm
\begin{abstract}
We systematically study three-terminal InSb-Al nanowire devices by using radio-frequency reflectometry. Tunneling spectroscopy measurements on both ends of the hybrid nanowires are performed while systematically varying the chemical potential, magnetic field and junction transparencies. Identifying the lowest-energy state allows for the construction of lowest- and zero-energy state diagrams, which show how the states evolve as a function of the aforementioned parameters. Importantly, comparing the diagrams taken for each end of the hybrids enables the identification of states which do not coexist simultaneously, ruling out a significant amount of the parameter space as candidates for a topological phase. Furthermore, altering junction transparencies filters out zero-energy states sensitive to a local gate potential. Such a measurement strategy significantly reduces the time necessary to identify a potential topological phase and minimizes the risk of falsely recognizing trivial bound states as Majorana zero modes.
\end{abstract}


\maketitle
\section{I. Introduction}
Superconductor-semiconductor hybrids have attracted great interest in recent years for their potential applications in creating Majorana zero modes (MZMs)  \cite{bib:Liang2008,bib:Roman2010,bib:Yuval2010}. Extensive experiments have been carried out on such hybrid nanowires \cite{bib:Vincent2012,bib:Mingtang2012,bib:Anindya2012,bib:Churchill2013,bib:Mingtang2016,bib:Chen2017,bib:Hao2021,bib:Vaitiekenas2020} and hybrid two-dimensional electron gases (2DEGs) \cite{bib:Fabriz2017,bib:Fornieri2019,bib:Ren2019}. Zero-bias peaks (ZBPs), observed at the ends of such hybrids, were initially considered as evidence for the existence of MZMs. However, such ZBPs could also originate from alternative trivial mechanisms, such as quasi-Majoranas \cite{bib:Adriaan2019}, disorder effects \cite{bib:JieLiu2012,bib:HainingPan2020,bib:HainingPan2021_1,bib:HainingPan2021_2,bib:HainingPan2021_3}, or a combination of Zeeman and Little-Parks effects \cite{bib:Valentini2021}. On the other hand, end-to-end correlations are a unique property of paired MZMs in a topological superconductor, and could be used to distinguish MZMs from trivial Andreev bound states in three-terminal architectures \cite{bib:Liu2013,bib:Lu2014,bib:Lai2019,bib:Anselmetti2019}.
Simulations taking into account the physical details of experimental devices (i.e.\ superconductor-semiconductor coupling, band offset at the interface, multiple subbands, and disorder effects) predict a significantly reduced and complex topological phase space \cite{bib:Antipov2018}. Therefore, finding such a phase in the large parameter space requires the development of a detection method capable of scanning the entire parameter space within a practical time \cite{bib:Pikulin2021}. Radio-frequency (rf) techniques have been succesfully implemented on superconducting qubits \cite{bib:Wallraff2004}, spin qubits \cite{bib:Hornibrook2014} and hybrid devices \cite{bib:Razmadze2019,bib:Jasper2019}. Compared to traditional dc conductance measurements, it enables a fast and high-resolution exploration of all essential parameters in hybrid devices.\\ 
     
Three-terminal InSb-Al nanowire devices are systematically investigated using rf reflectometry. Local tunneling spectroscopy is performed at two ends of the hybrid nanowires, while exploring the chemical potential (controlled by the so-called `super gate') and external magnetic field. The lowest-energy states (LESs) and zero-energy states (ZESs) are extracted as a function of super gate voltage and magnetic field, forming LES or ZES diagrams. As MZMs in an idealized model feature end-to-end correlations, the extracted diagrams of the two sides are compared to filter out uncorrelated ZESs. Stability of ZESs to transparency variation is studied by altering barrier gate settings, and zero-energy Andreev states residing around junctions are successfully identified. In addition, induced superconductivity on two ends of the hybrid nanowires is extracted, helpful for quantifing superconductor-semiconductor coupling in the hybrid nanowires. By applying the aforementioned experimental procedure, typical patterns of ZBPs are indentified in the studied devices, but after a closer inspection non-topological explanations are more likely. The approach is able to significantly accelerate the idenfication on a potential topological phase.    

\section{II. Experimental Setup}
\begin{figure*}[!t] 

\includegraphics[width=0.95\linewidth] {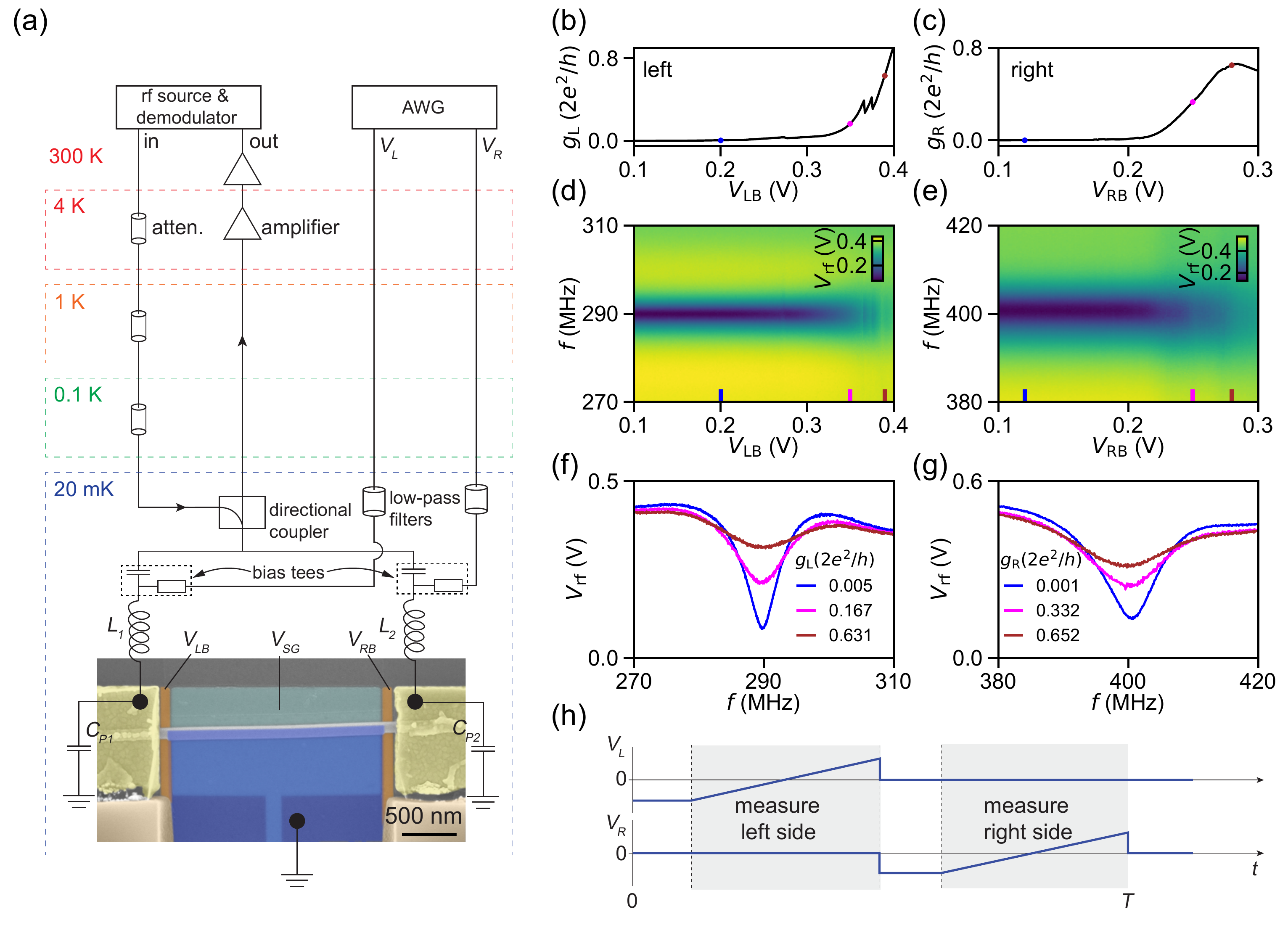}
\caption{Rf reflectometry measurement setup and its basic characteristics. \textbf{(a)} Measurement circuit together with a false-color SEM image of a three-terminal InSb-Al device. The superconducting lead (blue) is made from Al, connecting from the substrate to the nanowire. The superconductor-semiconductor hybrid is $\sim$2 $\mu$m long. Two probe leads (yellow), made from Ti/Au, are bonded to two superconducting inductors $L_{\mathrm{1}}$, $L_{\mathrm{2}}$. Voltages $V_{\mathrm{LB}}$ and $V_{\mathrm{RB}}$ are applied to bottom gates (orange) for tuning the left and right tunneling barriers, respectively. The chemical potential of the hybrid nanowire is tuned by the super gate (turquoise) with voltage $V_{\mathrm{SG}}$. \textbf{(b)} and \textbf{(c)} Dc conductance of left and right junction ($g_{\mathrm{L}}$, $g_{\mathrm{R}}$) versus corresponding barrier gate voltage ($V_{\mathrm{LB}}$, $V_{\mathrm{RB}}$). \textbf{(d)} and \textbf{(e)} Corresponding rf response as a function of barrier gates in the same range as \textbf{(b)} and \textbf{(c)}. \textbf{(f)} and \textbf{(g)} Line cuts at specific gate voltages from \textbf{(d)} and \textbf{(e)}. In \textbf{(d)-(g)}, $V_{\mathrm{rf}}$ is the amplitude of the reflected rf signal with $\sim$30 dB amplification at 4 K and $\sim$65 dB amplification at room temperature. \textbf{(h)} Bias-voltage waveforms applied at two probe leads as a function of time. In the shaded time period, voltage-bias spectroscopy of either side is performed. \textit{T} is the time period for measuring bias-spectroscopy traces on both sides.       
}\label{fig:1}
\end{figure*}

\begin{figure*}[!t] 
\centering
\includegraphics[width=0.95\linewidth] {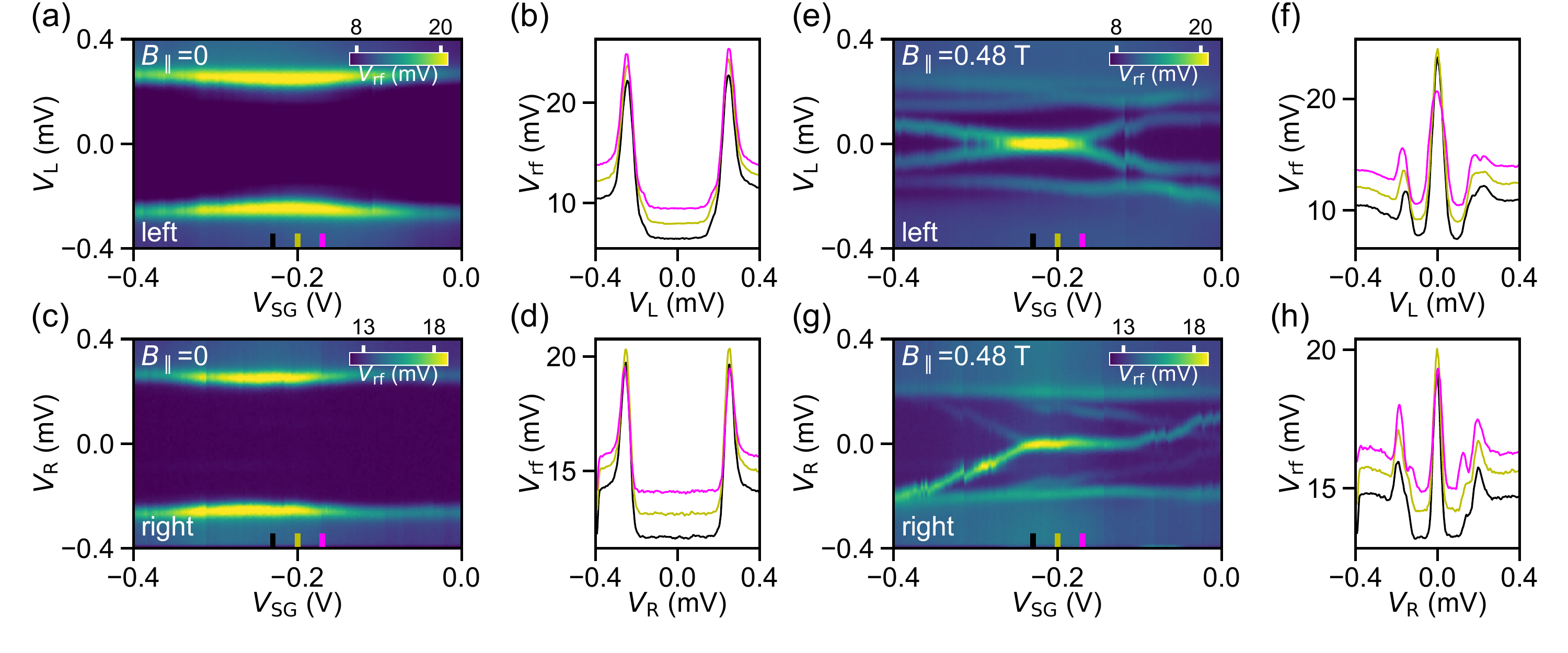}
\caption{\textbf{(a)} and \textbf{(e)} Selected segments of tunneling spectroscopy at the left junction at $B_{\mathrm{||}}$ = 0\,T and $B_{\mathrm{||}}$ = 0.48\,T, respectively. Line cuts are taken at colored bars and shown in \textbf{(b)} and \textbf{(f)}. \textbf{(c)} and \textbf{(g)} are similar as \textbf{(a)} and \textbf{(e)}, but for the right junction. \textbf{(d)} and \textbf{(h)} show line cuts from \textbf{(c)} and \textbf{(g)}. Line cuts at zero magnetic field show a hard gap on both sides. Results at $B_{\mathrm{||}}$ = 0.48\,T illustrate the coexistence of zero-bias peaks on two sides. Note that the curves shown in \textbf{(b)}, \textbf{(d)}, \textbf{(f)} and \textbf{(h)} are shifted vertically for better visibility.
}\label{fig:2}
\end{figure*}

Fig.~\ref{fig:1}(a) shows a circuit diagram of the measurement setup together with a false-color scanning electron microscope (SEM) image of Device 1. Three-terminal devices are fabricated from InSb nanowires \cite{bib:Gada2019} using the recently developed shadow-wall lithography technique, enabling high-quality semiconductor-superconductor quantum devices  \cite{bib:Sebastian2020,bib:Francesco2020}. In the SEM image, an Al film (blue) is connected from the substrate to the nanowire to serve as a superconducting drain lead, while Ti/Au contacts (yellow) are fabricated on both ends of the hybrid nanowire to serve as probe leads. Voltages $V_{\mathrm{LB}}$, $V_{\mathrm{RB}}$ are applied onto barrier gates (orange) to tune the transparency of the tunneling junctions. The voltage on the super gate (turquoise), $V_{\mathrm{SG}}$, is changing the chemical potential of the hybrid nanowire. In Fig. S1, we show SEM pictures of two additional measured devices and a schematic of the cross-section of the device. \\

In order to accelerate tunneling spectroscopy at both junctions, an rf-conductance measurement scheme is employed \cite{bib:Schoelkopf1998,bib:Reilly2007,bib:Razmadze2019}. The left and right probe lead of the device are connected to two superconducting spiral inductors ($L_{\mathrm{1}}$, $L_{\mathrm{2}}$) \cite{bib:Hornibrook2014}. Together with parasitic capacitors ($C_{\mathrm{P1}}$, $C_{\mathrm{P2}}$) to ground, the inductors form two rf resonant circuits with typical resonance frequencies 250-450 MHz (see optical images of inductor chips with devices in Fig. S2). Each resonator acts as an impedance transformer for the corresponding tunneling junction. On resonance, the typical junction impedance ($\sim$150\ k$\Omega$) is converted towards 50 $\Omega$, which is the characteristic impedance of the transmission lines in the cryostat. Consequently, the reflection of the rf circuits at resonance displays a sensitive dependence on the differential conductance of the tunneling junctions. Fig.~\ref{fig:1}(b) and Fig.~\ref{fig:1}(c) present the pinch-off curves of the junctions at the two ends of the hybrid nanowire, at 10 mV dc bias, while Fig.~\ref{fig:1}(d)-(g) show corresponding response of the resonator circuits. The rf reflection has a sensitive response to conductance changing from 0.005 $G_{\mathrm{0}}$ to 0.6\,$G_{\mathrm{0}}$ ($G_{\mathrm{0}}$=2$e^2/h$). Such a broad conductance response allows sensitive rf detection at different tunneling transparencies.\\ 

The integration time per data point is about 1 ms, approximately two orders of magnitude less than the integration time of a conventional lock-in conductance measurement. To take advantage of the reduced integration time, we employ a rastering scheme \cite{bib:Reilly2007,bib:Stehlik2015} to rapidly sweep the dc voltage bias applied at the tunneling junctions. Fig.~\ref{fig:1}(h) shows the waveforms generated by an arbitrary waveform generator (AWG). To perform the tunneling spectroscopy measurements at the two ends of the nanowire, the AWG generates a pair of triangular pulses, each sweeping the dc voltage bias at one of the junctions. The waveforms are accompanied by triggers, synchronizing the data aquisition with the voltage sweeps. Throughout the experiment, we perform pairs of tunneling spectroscopy measurement (typically 200 data points for each side with a total duration of $\sim$0.4 s), which we repeat while varying gate voltages and the external magnetic field.\\     
      
\section{III. Results}
\subsection{A. Tunneling spectroscopy}
Initially, basic characterization of devices and resonators is performed before moving to tunneling spectroscopy with rf. First, cross-talk between the super gate and barrier gates is measured at a fixed dc voltage bias. While sweeping $V_{\mathrm{SG}}$, barrier gate voltages are changed accordingly to maintain a constant junction conductance (see Fig. S3(a) and Fig. S3(b)). Next, the magnetic field is aligned along the nanowire axis. Furthermore, the resonator frequency shift in an external magnetic field is characterized (see Fig. S3(c) and Fig. S3(d)). As the external magnetic field is swept, the probing frequencies are adjusted to maintain a high sensitivity of the rf conductance measurement. Finally, tunneling spectroscopy is performed on both sides of the device by applying the dc bias waveform illustrated in Fig.~\ref{fig:1}(h), while stepping $V_{\mathrm{SG}}$ and the parallel magnetic field, $B_{\mathrm{||}}$.\\

\begin{figure}[!t]
\centering
\includegraphics[width=1\linewidth]{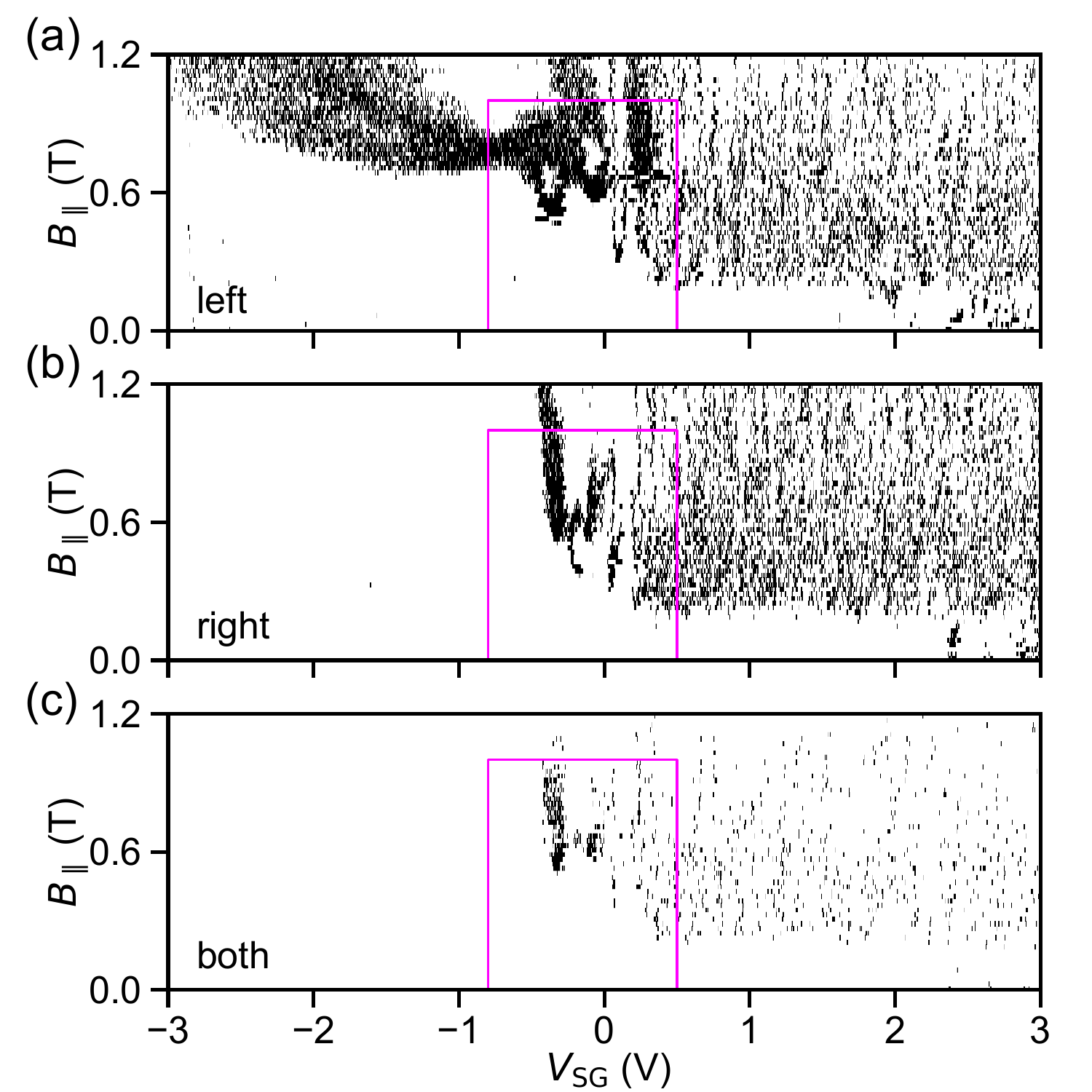}
\caption{ZES diagrams as a function of $V_{\mathrm{SG}}$ and $B_{\mathrm{||}}$ for the left junction \textbf{(a)} and right junction \textbf{(b)}.  Black and white pixels indicate the presence and absence of ZESs, respectively. \textbf{(c)} Diagram with black pixels indicating the coexistence of ZESs on both ends. In each panel, there are 2523 data points along $V_{\mathrm{SG}}$ and 61 along $B_{\mathrm{||}}$ (i.e.\ total number of pixels is 2523$\times$61). The full data set is acquired in $\sim$142 hours. Magenta rectangles mark the region with a relatively large density of coexistent ZESs.
}\label{fig:3}
\end{figure}

Fig.~\ref{fig:2} shows segments of tunneling spectroscopy measurements on two nanowire ends at different $B_{\mathrm{||}}$. Fig.~\ref{fig:2}(a) and Fig.~\ref{fig:2}(c) show the results as a function of $V_{\mathrm{SG}}$ at $B_{\mathrm{||}}$=0. Line cuts at different $V_{\mathrm{SG}}$ are presented in Fig.~\ref{fig:2}(b) and Fig.~\ref{fig:2}(d). Suppressed conductance in between two pronounced coherence peaks suggests a hard superconducting gap. The superconducting gap is $\sim$260 $\mu$eV, consistent with previous report based on the same fabrication platform \cite{bib:Sebastian2020,bib:Francesco2020}. Fig.~\ref{fig:2}(e) and Fig.~\ref{fig:2}(g) present an example of sub-gap features at $B_{\mathrm{||}}$=0.48\,T for the same $V_{\mathrm{SG}}$ range as Fig.~\ref{fig:2}(a) and Fig.~\ref{fig:2}(c). ZBPs are formed at both ends between $V_{\mathrm{SG}}$ = $\num{-0.23}$\,V and $V_{\mathrm{SG}}$ = $\num{-0.15}$\,V (see Fig.~\ref{fig:2}(f) and Fig.~\ref{fig:2}(h) for line cuts). ZBPs, peaks in differential conductance at zero energy, indicate the existence of ZESs, which are states with zero energy. Having established our setup for tunneling spectroscopy at both nanowire ends, we start mapping ZESs as a function of multiple parameters over a large range with high resolution.\\
 
\subsection{B. Diagrams of lowest- and zero-energy states}
For a given combination of $B_{\mathrm{||}}$ and $V_{\mathrm{SG}}$, the presence of ZBPs is validated by analyzing one $V_{\rm rf}$-$V_{\mathrm{L/R}}$ trace (see Fig. S5 for details). This is repeated for all measured parameter values and presented as ZES diagrams, as shown in Fig.~\ref{fig:3}. Fig.~\ref{fig:3}(c) shows coexisting ZESs on both ends of the hybrid nanowire. In these diagrams, three distinct regimes can be observed. (1) For negative super gate voltages ($V_{\mathrm{SG}}$ $<$ $\num{-0.5}$\,V), ZESs only appear at high magnetic field (on the order of $B_{\mathrm{||}}$ = 0.8\,T) and there are no coexistent ZESs. (2) For positive super gate voltages ($V_{\mathrm{SG}}$ $>$ 0.5\,V), ZESs are ubiquitous at fields as low as $B_{\mathrm{||}}$ = 0.2\,T. Here, coexistent ZESs are sparsely distributed in parameter space. (3) In an intermediate regime ($\num{-0.5}$\,V $<$ $V_{\mathrm{SG}}$ $<$ $\num{0.5}$\,V), ZESs emerge at moderate magnetic fields compared to the other two regimes. Notably, ZESs form regular shapes in parameter space and there is a significant amount of coexisting ZESs. This behavior is reproduced for two other InSb-Al hybrid nanowires presented in this work (see Fig. S11 and Fig. S13). A recent work on InSb-Al hybrid islands reports three similar regimes in $V_{\mathrm{SG}}$ \cite{bib:Jie2020}. It is explained by a tunable superconductor-semiconductor coupling with $V_{\mathrm{SG}}$ \cite{bib:Antipov2018,bib:Mikkelsen2018}.  The intermediate regime is identified to be the most promising region to search for a topological superconducting phase. We focus the subsequent measurements on this intermediate super gate regime (marked by magenta rectangles in Fig.~\ref{fig:3}).\\ 

\begin{figure*}[!t]
\centering
\includegraphics[width=0.97\linewidth]{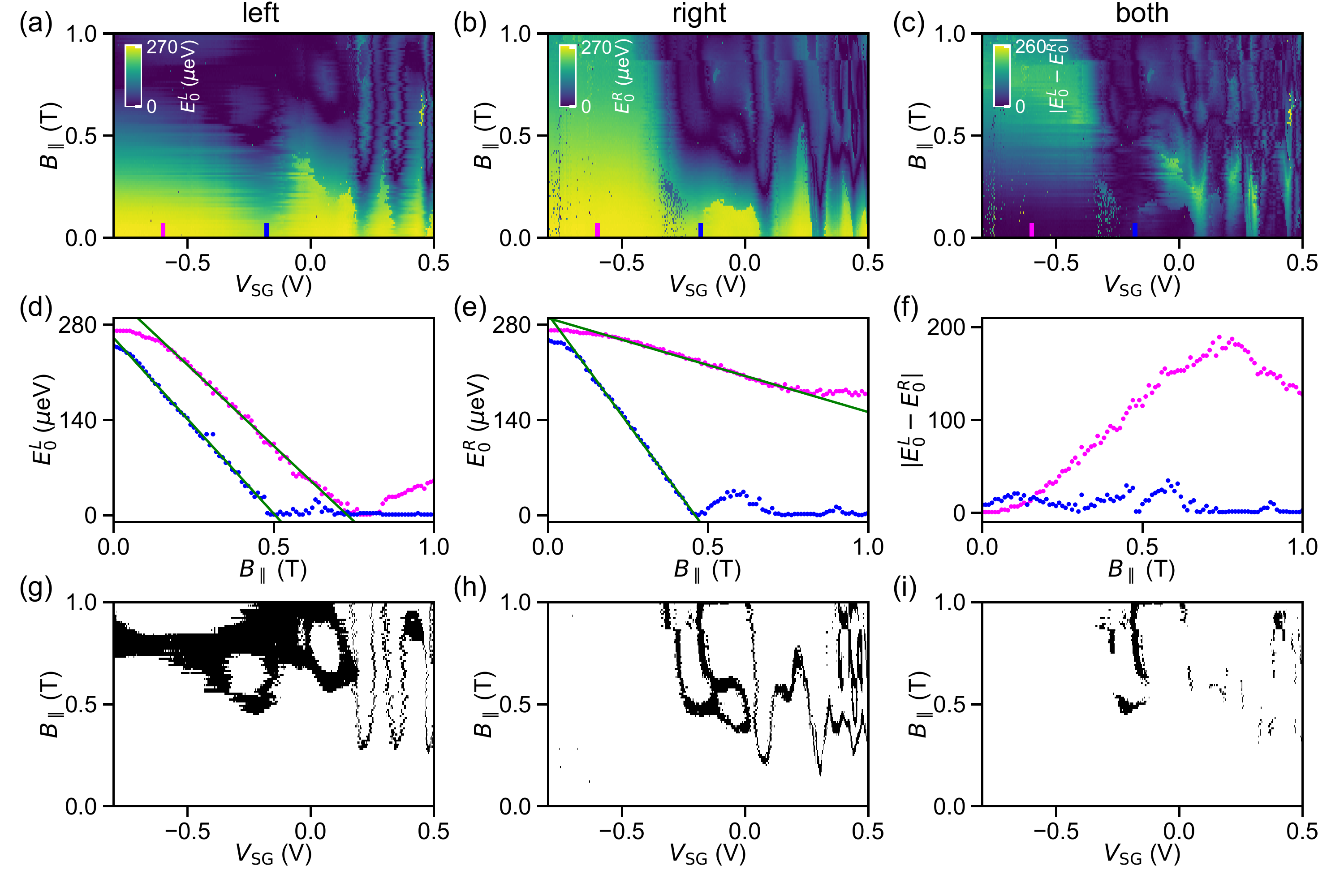}
\caption{High-resolution LES and ZES diagrams. \textbf{(a)} Energy of LESs, $E_{\mathrm{0}}^{\mathrm{L}}$, probed on the left junction as a function of $V_{\mathrm{SG}}$ and $B_{||}$. \textbf{(b)} Energy of LESs, $E_{\mathrm{0}}^{\mathrm{R}}$, probed on the right junction versus $V_{\mathrm{SG}}$ and $B_{\mathrm{||}}$. \textbf{(c)} Energy difference between the LESs probed on the two ends. \textbf{(d)-(f)} Line cuts taken at different $V_{SG}$ from \textbf{(a)-(c)}. In \textbf{(d)} and \textbf{(e)}, solid green lines are linear fits. \textbf{(g)} and \textbf{(h)} ZES diagrams for the left and right junction, respectively. \textbf{(i)} Diagram of coexisting ZESs on both ends. Gates are swept by following the cyan dashed lines ($g_{\mathrm{L}}$~$ \sim $0.035\,G$_{\mathrm{0}}$, $g_{\mathrm{R}}$~$ \sim $0.064\,$G_{\mathrm{0}}$) in Fig. S4. In \textbf{(a)-(c)} and \textbf{(g)-(i)}, each panel includes 522$\times$101 pixels. The dataset is acquired in $\sim$35 hours.
}\label{fig:4}
\end{figure*}

Fig.~\ref{fig:4} presents high-resolution diagrams obtained in the intermediate super gate regime. Fig.~\ref{fig:4}(a) and Fig.~\ref{fig:4}(b) show the energy of LESs probed at the two junctions, \textit{E$_{0}^{L}$} and \textit{E$_{0}^{R}$}, versus $V_{\mathrm{SG}}$ and $B_{\mathrm{||}}$ (LES extraction method is shown in Fig. S5). At zero magnetic field, $E_{\mathrm{0}}^{\mathrm{L/R}}$ is close to the superconducting gap ($\sim$260 $\mu$eV). As $B_{\mathrm{||}}$ increases, $E_{\mathrm{0}}^{\mathrm{L/R}}$ starts to drop due to the emergence of sub-gap states. In order to illustrate the dependence of $E_{\mathrm{0}}^{\mathrm{L/R}}$ on $B_{\mathrm{||}}$, examples of two vertical line cuts are shown in Fig.~\ref{fig:4}(d) and Fig.~\ref{fig:4}(e). These are picked to illustrate two types of behavior: For the blue line cuts ($V_{\mathrm{SG}}$ = $\num{-0.18}$\,V), the behavior on both sides of the nanowire is similar. Sub-gap states emerge and drop to zero energy, with a comparable effective g-factor (solid green lines are fitting traces to the linear part of the data). On the other hand, the magenta line cuts ($V_{\mathrm{SG}}$ = $\num{-0.6}$\,V) show an example where on one junction, a sub-gap state drops to zero energy while on the other side no sub-gap states emerge. In order to identity LESs that may extend between the two ends of the hybrid nanowire, the energy difference between LESs, $|{E_{\mathrm{0}}^{\mathrm{L}}}-E_{\mathrm{0}}^{\mathrm{R}}|$, is calculated and shown in Fig.~\ref{fig:4}(c). Fig.~\ref{fig:4}(f) shows the line cuts at the same $V_{\mathrm{SG}}$ as in Fig.~\ref{fig:4}(d) and Fig.~\ref{fig:4}(e). Notably, for the blue line cut the energy difference of the LESs on both ends is close to zero within a large range of field, indicating a potential correlation. In contrast, the magenta line cut shows a large energy difference for almost all field values, which signifies uncorrelated behavior.\\  

From the LES diagrams, states with zero energy are identified. These states are presented in ZES diagrams, shown in  Fig.~\ref{fig:4}(g)-Fig.~\ref{fig:4}(i) (see the example of the extraction process in Fig. S5). Similar to Fig.~\ref{fig:3}, black pixels in Fig.~\ref{fig:4}(g) and Fig.~\ref{fig:4}(h) represent the presence of ZESs, while white pixels indicate the absence of ZESs. Regular features are observed in these diagrams, including parabolic and oscillatory shapes, as well as clusters of black pixels. The intersection of the two diagrams yields a diagram (Fig.~\ref{fig:4}(i)) consisting of ZESs which coexist on both sides. Around $V_{\mathrm{SG}}$ $\sim$ $\num{-0.2}$\,V and $B_{\mathrm{||}}$ $>$ 0.5\,T, a high density of coexistent ZESs is observed which indicates a potential candidate region for a topological phase. In the following section, this region and several of the regular patterns will be further analyzed.
              
\subsection{C. Detailed analysis of ZES and LES diagrams}

\subsubsection{C1. Coexisting ZBP clusters}

\begin{figure*}[!t]
\centering
\includegraphics[width=0.90\linewidth]{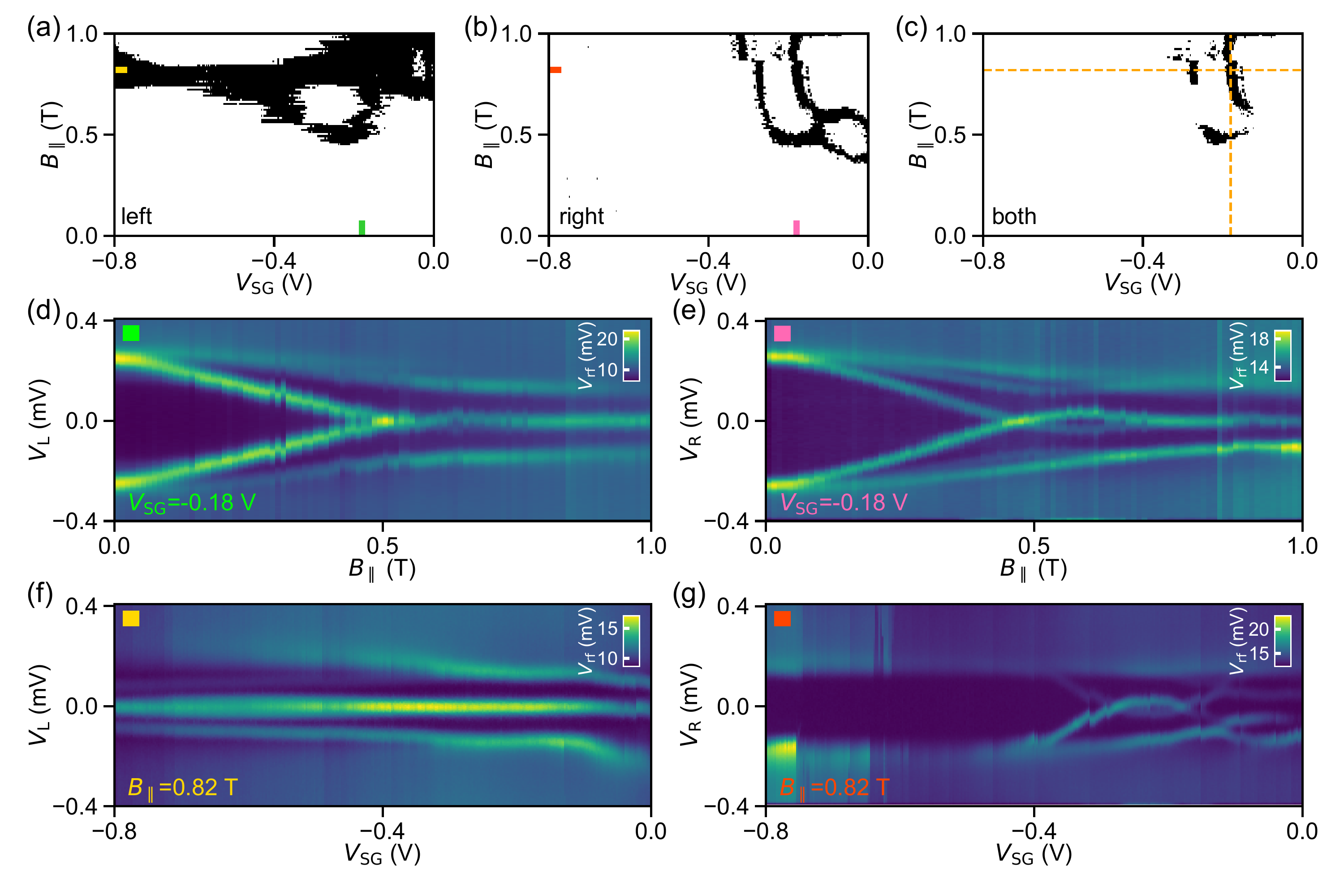}
\caption{Clusters of ZBPs with respect to super gate and magnetic field. \textbf{(a)}-\textbf{(c)} ZES diagrams for the left junction, the right junction, and coexistence in both junctions (same as Fig.~\ref{fig:4}(g)-Fig.~\ref{fig:4}(i), but with reduced super gate range). In \textbf{(a)}, the green mark indicates the position of $V_{\mathrm{SG}}$ = $\num{-0.18}$\,V and corresponding tunneling spectroscopy data is shown in \textbf{(d)}. The results at $B_{\mathrm{||}}$ = 0.82\,T (yellow mark) is shown in \textbf{(f)}. In \textbf{(b)}, tunneling spectroscopy data of the vertical (pink mark) and horizontal (red mark) line cut is shown in \textbf{(e)} and \textbf{(g)}, respectively. In \textbf{(c)}, the horizontal and vertical dashed line indicates $B_{\mathrm{||}}$ = $\num{0.82}$\,T and $V_{\mathrm{SG}}$ = $\num{-0.18}$\,V, respectively.   
}\label{fig:5}
\end{figure*}
ZES diagrams constructed in the previous section identify a region with coexisting ZESs on both ends of the nanowire hybrid. Fig.~\ref{fig:5}(a)-Fig.~\ref{fig:5}(c) show a zoom-in of Fig.~\ref{fig:4}(g)-Fig.~\ref{fig:4}(i). At fixed $V_{\mathrm{SG}}$ = $\num{-0.18}$\,V, tunneling spectroscopy on both ends of the hybrid nanowire shows a sub-gap state reaching zero energy around 0.5\,T (Fig.~\ref{fig:5}(d) and Fig.~\ref{fig:5}(e)). After a single oscillation, it sticks to zero energy for $\sim$300\,mT. On the other hand, when the magnetic field is fixed at 0.82\,T, a stable ZES is observed on the left side (Fig.~\ref{fig:5}(f)). Remarkably, it persists for more than 800\,mV in $V_{\mathrm{SG}}$. Considering that the dielectric layer is made from 20\,nm HfO$_{x}$, chemical potential is changed by a significant amount in this super gate range. Such stable ZBPs as a function of super gate voltage and magnetic field, together with similar behavior on both sides of the device as a function of magnetic field, have been previously interpreted as evidence of MZMs. However, tunneling spectroscopy along super gate on the right side of the device (Fig.~\ref{fig:5}(g)) reveals a strikingly different behavior, with only two crossings through zero energy. The different behaviors on the two sides of the device can be recognized as well from ZES diagrams in Fig.~\ref{fig:5}(a)-Fig.~\ref{fig:5}(c). Such clearly distinct behavior with respect to changes in the chemical potential implies that the ZESs on the two sides of the device do not originate from an unbroken topological superconducting phase.\\

\subsubsection{C2. Parabolic patterns in ZES diagrams}
\begin{figure}[!t]
\centering
\includegraphics[width=1.0\linewidth]{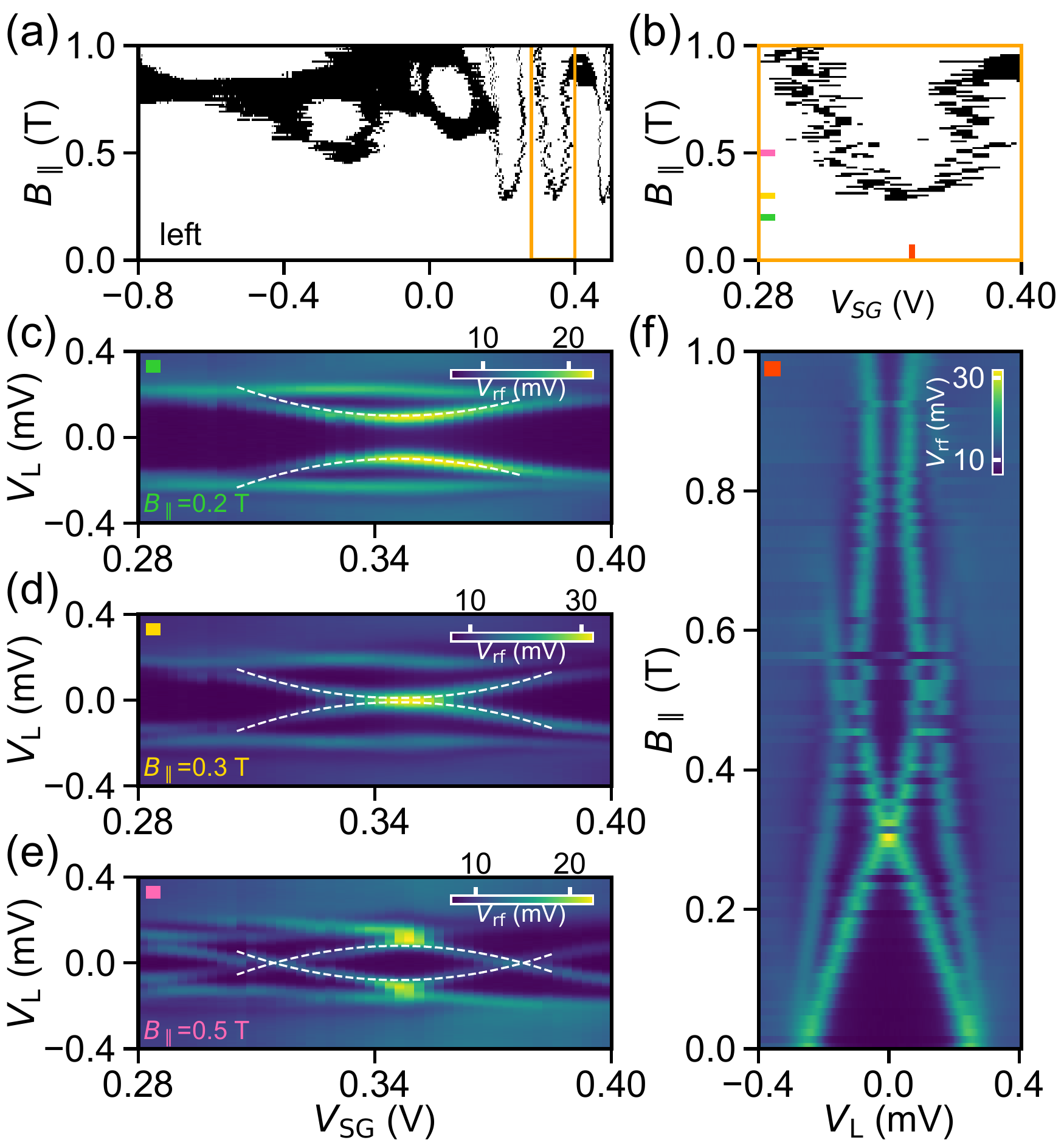}
\caption{Parabolic patterns in ZES diagrams. \textbf{(a)} ZES diagram for the left junction (same as Fig.~\ref{fig:4}(g)). The region marked by the orange rectangle is re-plotted in \textbf{(b)}. Three line cuts are made at different $B_{\mathrm{||}}$ and corresponding tunneling spectroscopy data is displayed in \textbf{(c)-(e)}. White dashed lines serve as guide to the eye. \textbf{(f)} Evolution of sub-gap levels in magnetic field at $V_{\mathrm{SG}}$ = 0.35\,V (red vertical line cut from \textbf{(b)}). 
}\label{fig:6}
\end{figure}

In Fig.~\ref{fig:4}, ZESs form parabolic patterns in the $V_{\mathrm{SG}}$-$B_{\mathrm{||}}$ space. Such parabolic patterns can represent the onset of a topological phase when Majorana zero modes at two ends of a short hybrid nanowire strongly interact \cite{bib:Sarma2012,bib:Chen2017}. In Fig.~\ref{fig:6}, an example of such a parabola is shown and its tunneling spectroscopy data is presented. In Fig.~\ref{fig:6}(a), an orange rectangle marks the region with such a parabolic pattern. This region is re-plotted in Fig.~\ref{fig:6}(b). In order to understand the pattern, three line cuts are made at different magnetic fields and corresponding spectroscopic results are plotted in Fig.~\ref{fig:6}(c)-Fig.~\ref{fig:6}(e). At $B_{\mathrm{||}}$ = 0.2\,T (Fig.~\ref{fig:6}(c)), there is a pair of levels emerging below the superconducting gap, marked by white dashed lines. Once the magnetic field is increased to 0.3\,T (Fig.~\ref{fig:6}(d)), the pair of sub-gap levels merges at zero energy, forming ZBPs. This field corresponds to the onset of ZESs in Fig.~\ref{fig:6}(b). At higher $B_{\mathrm{||}}$, for example $B_{\mathrm{||}}$ = 0.5\,T (Fig.~\ref{fig:6}(e)), the sub-gap levels form two crossings at zero energy. The evolution of these sub-gap levels in $B_{\mathrm{||}}$ at $V_{\mathrm{SG}}$ = 0.35\,V is shown in Fig.~\ref{fig:6}(f). In Fig. S6, another parabolic pattern which has similar properties as Fig.~\ref{fig:6} is presented. This behavior is fully explained by Zeeman-driven Andreev level splitting in a quantum dot proximitized by a superconducting lead \cite{bib:Lee2014}.\\ 

\subsubsection{C3. Oscillatory patterns in LES diagrams}
\begin{figure*}[!t]
\centering
\includegraphics[width=0.90\linewidth]{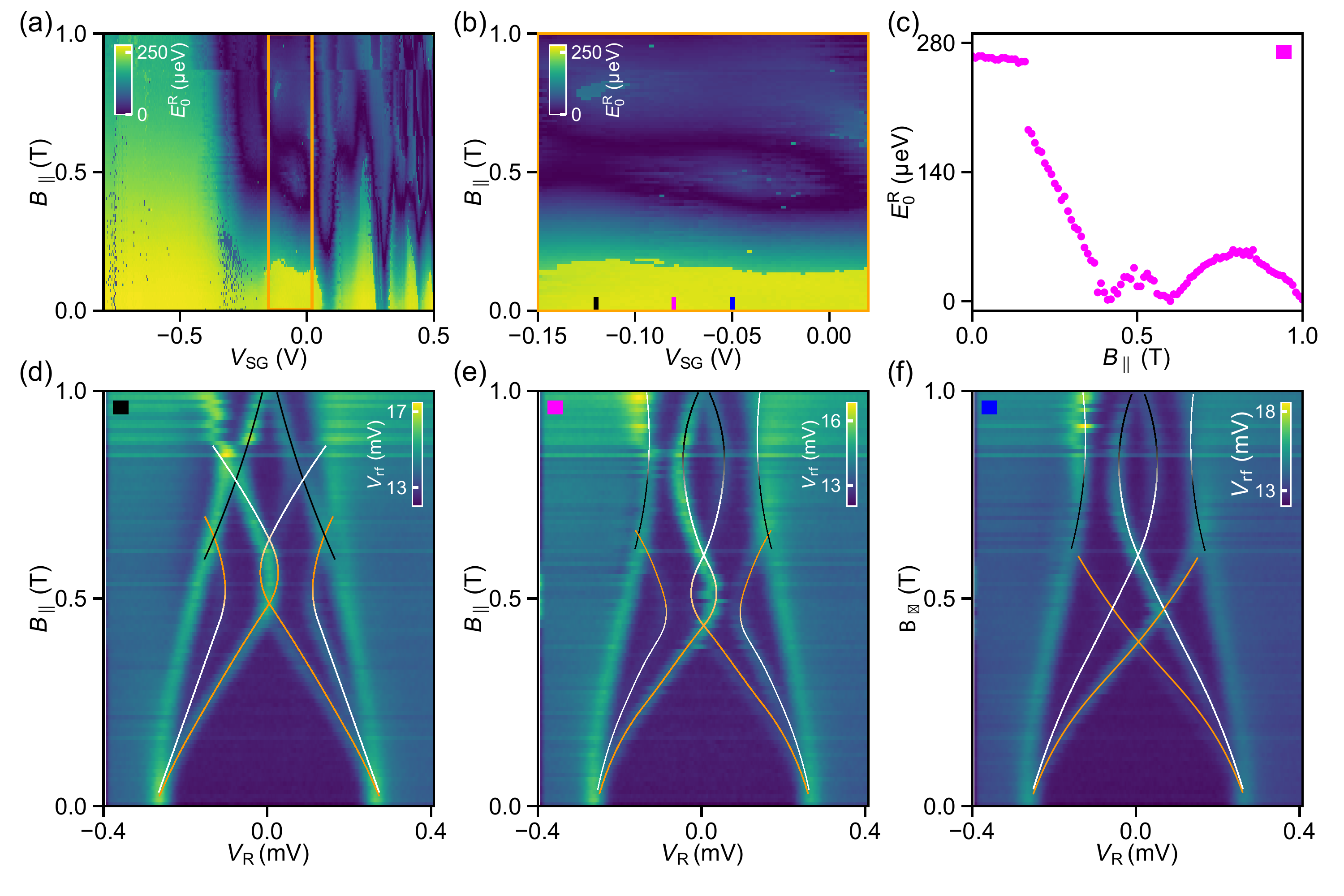}
\caption{Oscillatory patterns in LES diagrams. \textbf{(a)} LES diagram for the right junction (same as Fig.~\ref{fig:4}(b)). The region marked by the orange rectangle is plotted in \textbf{(b)}. \textbf{(c)} Example of an oscillating LES, showing $E_{\mathrm{0}}^{\mathrm{R}}$ versus $B_{\mathrm{||}}$ for the line cut taken at $V_{\mathrm{SG}}$ = $\num{-0.08}$\,V (marked by the magenta bar in \textbf{(b)}). Tunneling spectroscopy data taken at $V_{\mathrm{SG}}$ = $\num{-0.12}$\,V (black bar), $V_{\mathrm{SG}}$ = $\num{-0.08}$\,V (magenta bar), and $V_{\mathrm{SG}}$ = $\num{-0.05}$\,V (blue bar) is shown in \textbf{(d)}, \textbf{(e)} and \textbf{(f)}, respectively. In \textbf{(d)-(f)}, three sub-gap states are marked by orange, white, and black lines which serve as a guide to the eye. Gradient colors indicate interaction between states.  
}\label{fig:7}
\end{figure*}

The oscillation of LESs in magnetic field with an increasing amplitude and period is consistent with the prediction of smoking gun evidence for MZMs \cite{bib:Sarma2012}. This type of behavior would result in oscillatory patterns in LES and ZES diagrams. An example of such patterns is shown in Fig.~\ref{fig:7}(a) and Fig.~\ref{fig:7}(b). A line cut through such a circle, shown in Fig.~\ref{fig:7}(c) ($V_{\mathrm{SG}}$ = $\num{-0.08}$\,V), shows the energy of a LES dropping to zero before oscillating with increasing amplitude and period, thus matching the smoking gun predictions. The tunneling spectroscopy data as a function of $B_{\mathrm{||}}$ corresponding to this particular line cut is shown in Fig.~\ref{fig:7}(e). In this map, three discrete sub-gap states are observed and marked with an orange, white, and black lines. The LES (orange) first comes down in energy and crosses through zero energy, before interacting with another state (white). Interaction between these states results in an anti-crossing, which can be attributed to the mixing of different spin species of two states via spin-orbit interaction \cite{bib:Moor2018}, and is represented by a gradient color. The LES crosses through zero once more, and subsequently interacts with another state (black). In addition, two spectroscopy results at a lower ($V_{\mathrm{SG}}$ = $\num{-0.12}$\,V) and higher ($V_{\mathrm{SG}}$ = $\num{-0.05}$\,V) super gate voltage are presented in Fig.~\ref{fig:7}(d) and Fig.~\ref{fig:7}(f), respectively. By changing the super gate voltage, the magnitude of the interaction between the states can be tuned. Consequently, the interactions can become negligible which results in states crossing rather than anti-crossing. This indicates that the results in Fig.~\ref{fig:7}(e) or Fig.~\ref{fig:7}(c) arise from several sub-gap states interacting with each other. Thus, while the behavior of LESs in Fig.~\ref{fig:7}(e) and Fig.~\ref{fig:7}(c) is consistent with the predicted evidence of MZMs, such oscillations can originate from anti-crossings between the LES and other states. The analysis made above suggests that although oscillatory patterns in LES and ZES diagrams are expected for interacting MZMs, they may also originate from interactions between topologically trivial Andreev bound states.\\

\begin{figure}[hp] 
\centering
\includegraphics[width=0.97\linewidth] {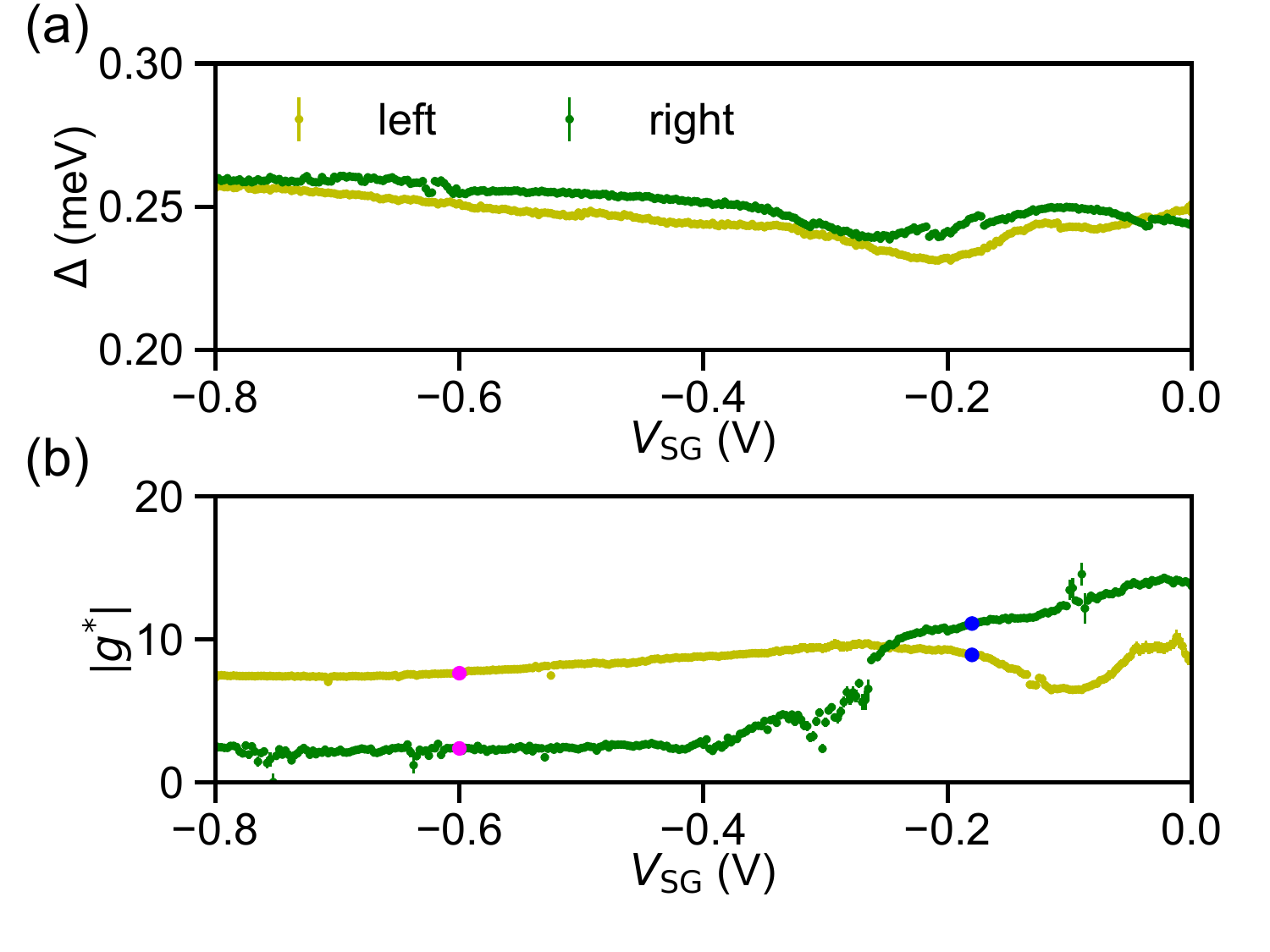}
\caption{Gate-dependent induced superconductivity. \textbf{(a)} Superconducting gap as a function of super gate voltage \textit{V$_{SG}$} extracted from the left (yellow) and right (green) junction. \textbf{(b)} Effective g-factor versus \textit{V$_{SG}$} for the left (yellow) and right (green) end of the hybrid nanowire. Magenta and blue points correspond to the data extracted from the traces in Fig.~\ref{fig:4}(d) and Fig.~\ref{fig:4}(e).    
}\label{fig:8}
\end{figure}

\subsubsection{C4. Induced superconductivity}
In addition to extracting of the energy of various sub-gap states in the system, two other important parameters can be determined from the data: the superconducting gap size $\Delta$ and the effective g-factor $g^{*}$. From the local tunneling spectroscopy measurements, the superconducting gap can be estimated by fitting zero-field \textit{g$_{L/R}$}-\textit{V$_{L/R}$} traces with the BCS$-$Dynes formula \cite{bib:Dynes1978}, where \textit{g$_{L/R}$} is obtained from \textit{V$_{rf}$}-\textit{g$_{L/R}$} correspondence in Fig.~\ref{fig:1}. On the other hand, the effective g-factor is determined by making a linear fit to the energy dependence of the LES as a function of magnetic field. In Fig.~\ref{fig:8}, the evolution of these two parameters is shown as a function of super gate voltage. At both ends of the nanowire, the estimated gap size behaves similarly and remains largely unaffected by the super gate. It only shows a small dip in the vicinity of \textit{V$_{SG}$}=$\num{-0.25}$\,V, and from the spectra shown in Fig.~\ref{fig:2} it can be seen that these dips correspond to the energy minima of two LESs. However, if we instead look at the field evolution it becomes apparent that the extracted g-factors do not behave in a similar way. This indicates that the LESs at two ends are uncorrelated, which offers an alternative perspective to the energy evolution which we investigated in previous sections. In addition, the right side of the device shows an absence of sub-gap states below \textit{V$_{SG}$}=$\num{-0.4}$\,V (see an example in Fig. S7). The corresponding g-factor is estimated from the gap edge and remains close to 2, as the measured properties are dominated by the Al film. It is worth to note that with local tunneling spectroscopy, only the superconducting properties in the vicinity of the junctions are detected while leaving the bulk properties inaccessible. Thus, these measurements could be complemented by non-local measurements in order to investigate the induced superconductivity  in the bulk of the hybrids \cite{bib:Pikulin2021}.

\subsection{D. Influence of barrier gates}
 
\begin{figure*}[!t] 
\centering
\includegraphics[width=0.97\linewidth] {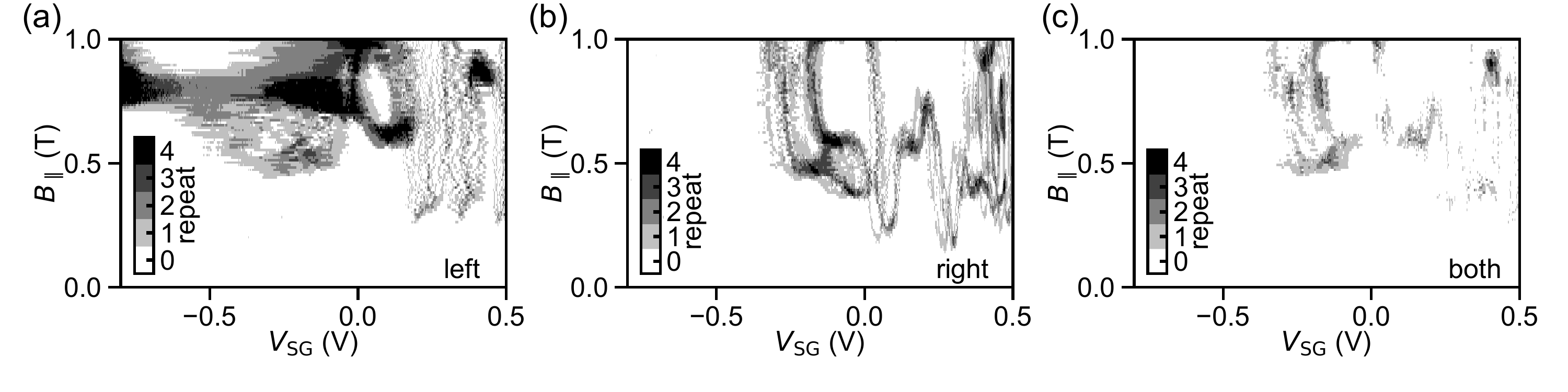}
\caption{Overlapping ZES diagrams taken at different barrier gate settings for the left junction \textbf{(a)}, right junction \textbf{(b)}, and coexistence in both junctions \textbf{(c)}. The panels are obtained by overlapping ZES diagrams from Fig. S8 ($g_{\mathrm{L}}\sim$\,0.027\,$G_{\mathrm{0}}$, $g_{\mathrm{R}}\sim$\,0.025\,$G_{\mathrm{0}}$), Fig. S9 ($g_{\mathrm{L}}\sim$\,0.029\,$G_{\mathrm{0}}$, $g_{\mathrm{R}}\sim$\,0.004\,$G_{\mathrm{0}}$), Fig.~\ref{fig:4} ($g_{\mathrm{L}}\sim$\,0.03\,$G_{\mathrm{0}}$, $g_{\mathrm{R}}\sim$\,0.064\,$G_{\mathrm{0}}$), and Fig. S10 ($g_{\mathrm{L}}\sim$\,0.047\,$G_{\mathrm{0}}$, $g_{\mathrm{R}}\sim$\,0.129\,$G_{\mathrm{0}}$).   
}\label{fig:9}
\end{figure*}

ZESs formed in the vicinity of the junction can mimic MZM behavior with respect to magnetic field and chemical potential variations \cite{bib:ChunXiao2017}. Tunnel transparency is an important experimental parameter which can be used for distinguishing MZMs from trivial Andreev bound states. To investigate the stability of ZESs shown in Fig.~\ref{fig:4}, similar measurements are performed with three different barrier gate settings. The corresponding processed diagrams are plotted in Fig. S8, Fig. S9, and Fig. S10 in the supplementary material. ZES diagrams obtained for four different barrier gate settings are overlapped to form ZES histograms in Fig.~\ref{fig:9}. As shown in Fig.~\ref{fig:9}(a), the left end of the hybrid exhibits clusters of ZESs in the space of chemical potential and magnetic field. As those ZESs are stable against variation of all experimentally accessible parameters, they are compatible with MZMs. However, as shown in Fig.~\ref{fig:9}(b), the right side of the studied hybrid does not show similar stability and thus indicates that the device does not exhibit an unbroken topological superconducting phase. As expected, the uncorrelated behavior between left and right end of the hybrid is recognized as well in the histogram of coexisting ZESs (Fig.~\ref{fig:9}(c)). Similar studies have been performed on the other two devices and details are presented in Fig. S11-Fig. S14 in the supplementary material. The results do not yield evidence for correlated MZMs either.

\section{IV. Discussion}
In this work, three-terminal InSb-Al hybrid nanowire devices have been systematically studied using rf reflectometry. This approach introduces a critical experimental technique to quickly map out a large phase space at both ends of superconductor-semiconductor hybrid nanowires, which is crucial for searching for candidate regions of a topological phase. A wide range of chemical potential can be mapped out in high resolution by varying the super gate voltage, and consequently the chance of missing topologically non-trivial regions is minimized. Tunneling spectroscopy depending on super gate voltage and magnetic field enables the extraction of LESs  and ZESs, as well as induced superconducting gap and effective g-factor. Constructed diagrams of ZESs and LESs indicate the most promising regions for searching for MZMs \cite{bib:Pikulin2021}. Combined with the tunneling-spectroscopy data, clusters of ZPBs, parabolic and oscillatory patterns in ZES and LES diagrams are analyzed for the first time in large scale and high resolution. While such patterns mimic the predicted behavior of MZMs, the systematic exploration of all accessible experimental parameters suggest a non-topological origin. Further analysis by altering barrier gates indicate that the ZESs and LESs observed in this work are likely localized in the vicinity of the tunnel junctions. Additionally, the simultaneous detection of LESs at both ends of the hybrid enables the possibility to look for correlated ZESs, which is essential to prove the existence of paired MZMs. From the comparison of the LES and ZES diagrams at two ends of the hybrid, no indication of correlated behavior is observed in this work.\\
 
The absence of an unbroken topological superconducting phase in these samples can be attributed to several physical origins. Possible reasons include disorder, inhomogeneous interface band bending, local chemical potential fluctuations and a non-perfect detection method. Here, four possible shortcomings are elaborated. (1) In addition to forming trivial sub-gap levels which can mimic MZMs, disorder effects could push the Majorana wavefunctions away from the ends of the hybrid nanowires \cite{bib:Liu2013,bib:HainingPan2021_3}. In this case, even if a topological phase is formed in the bulk, the MZM wavefunction cannot be probed by tunneling probes at the edge of the hybrid nanowire. (2) Local chemical potential fluctuations with an energy above the helical gap can break a single topological phase into segments. In this case, the two ends of a hybrid nanowire are not necessarily correlated. In real devices, such fluctuations can originate from non-uniform gating effects, grain boundaries in aluminium, and disorder from impurities. (3) Recently, the interface band bending between the superconductor and semiconductor was recognized as an important ingredient for tuning the properties of the hybrid \cite{bib:Antipov2018,bib:Mikkelsen2018}. In particular, band bending can lead to the occupation of multiple subbands, and for the InSb-Al system the experimental implications are still unknown. This makes it difficult to predict the experimental conditions to achieve a topological superconducting phase. (4) The experimental protocol in this work is specifically designed to search for paired MZMs at two ends of an extended topological phase. If for any reason, a topological phase would not be continuous across the entire hybrid, it is possible that MZMs can be probed on one end but not on the other. In this case, the protocol applied in this work would miss a potential topological phase in the parameter space. Such a false negative cannot be ruled out by these measurements.\\

For the aforementioned problems, several solutions can be proposed. (1) The InSb nanowires in this work have transport mobilities of $\sim$\,4\,$\times$\,10$^{\mathrm{4}}$\,cm$^2$/V$\cdot$s \cite{bib:Gada2019}. The introduction of capping layers on top of the semiconductor could alleviate disorder and improve transport mobilities \cite{bib:Shabani2016}. (2) In order to improve resilience against local chemical potential fluctuations, alternative superconductors with larger superconducting gaps can be considered. Recent reports have successfully realized growth of Sn ($\Delta \sim$\,700\,$\mu$eV) \cite{bib:Pendharkar2021} and Pb ($\Delta \sim$\,1.25\,meV) \cite{bib:Kanne2021} on semiconductor nanowires. These hybrids may have a higher chance to achieve an uninterrupted topological phase in spite of the challenges in fabrication. (3) Band bending at the interface between superconductors and semiconductors can be engineered by inserting modulation layers  \cite{bib:Shabani2016}. Simultaneously, proper engineering of these layers can also be used as a tool to influence the magnitude of the induced superconducting gap. (4) Non-local measurements, proposed to probe the superconducting gap on three-terminal devices \cite{bib:Tomas2018}, can complement the fast rf reflectometry used in this work \cite{bib:Pikulin2021}. Corresponding experimental work established its capability of detecting bulk properties beyond the local characteristics \cite{bib:Gerbold2020,bib:D2020}, though the measurement speed was slow. Combining our present protocol with non-local measurement would strike a balance between measurement speed and detection reliability.\\
The experimental protocol developed in this work, together with possible improvements discussed above, will pave the way for unambiguously detecting MZMs in superconductor-semiconductor hybrid systems in the future. \\         
           
\section{Acknowledgement.} 
We thank Michael Wimmer and Anton Akhmerov for valuable comments on our manuscript. We also thank John M. Hornibrook and David J. Reilly providing the frequency multiplexing chips. We are grateful to Raymond Schouten, Olaf Benningshof and J. Mensingh for valuable technical support. This work has been financially supported by the Dutch Organization for Scientific Research (NWO) and Microsoft Corporation Station Q. \\     

\section{Author contribution.} 
J.-Y.W., S.H. and L.P.K. conceived the experiment. J.-Y.W., G.M., N.v.L., M.L., F.B. and M.Q.P. fabricated the devices. J.-Y.W., V.L. and F.M. performed the measurements. S.G., G.B. and E.P.A.M.B. carried out the material growth. J.-Y.W. analyzed the data. J.-Y.W., N.v.L., G.P.M, V.L., and L.P.K wrote manuscript with comments from all authors. J.-Y.W., S.H. and L.P.K. supervised the project.\\

\end{document}


\hsize\textwidth\columnwidth\hsize\csname@twocolumnfalse\endcsname

\title{Supplementary Material: Parametric exploration of zero-energy modes in three-terminal InSb-Al nanowire devices}
\author{Ji-Yin Wang}
\email{wangjiyinshu@gmail.com}
\address{QuTech and Kavli Institute of Nanoscience, Delft University of Technology, 2600 GA Delft, The Netherlands}

\author{Nick van Loo}
\author{Grzegorz P. Mazur}
\author{Vukan Levajac}
\author{Filip K. Malinowski}
\author{Mathilde Lemang}
\author{Francesco Borsoi}

\address{QuTech and Kavli Institute of Nanoscience, Delft University of Technology, 2600 GA Delft, The Netherlands}

\author{Ghada Badawy}
\author{Sasa Gazibegovic}
\address{Department of Applied Physics, Eindhoven University of Technology, 5600 MB Eindhoven, The Netherlands}

\author{Erik P.A.M. Bakkers}
\address{Department of Applied Physics, Eindhoven University of Technology, 5600 MB Eindhoven, The Netherlands}

\author{Marina Quintero-Pérez}
\address{QuTech and Kavli Institute of Nanoscience, Delft University of Technology, 2600 GA Delft, The Netherlands}
\address{Netherlands Organisation for Applied Scientific Research (TNO), Delft 2600 AD, Netherlands}

\author{Sebastian Heedt}
\address{QuTech and Kavli Institute of Nanoscience, Delft University of Technology, 2600 GA Delft, The Netherlands}

\author{Leo P. Kouwenhoven}
\address{QuTech and Kavli Institute of Nanoscience, Delft University of Technology, 2600 GA Delft, The Netherlands}

\date{\today}

\vskip1.5truecm
\maketitle
\section{Methodes}

\noindent\textbf{Device fabrication.} The InSb nanowires, with a typical diameter of 100\,nm, are grown on InSb (111)B substrates covered with pre-patterned SiN$_{x}$ mask via metalorganic vapour-phase epitaxy (MOVPE) \cite{bib:Badawy2019}. InSb nanowires are transfered onto pre-patterned substrates with a micro-manipulator. Hydrogen cleaning is used to remove the native oxide on the nanowire surface. Subsequently, a thin aluminum film ($\sim$\,15\,nm) is evaporated at 138\,K and a 30 degree angle with respect to the substrate using shadow-wall lithography  \cite{bib:Sebastian2020}. Finally, e-beam evaporation is used to make Ti/Au (10/120\,nm) contacts on the nanowire ends right after the removal of the NW oxide using argon ion milling.\\

\noindent\textbf{Transport measurements.} Samples are measured at a base temperature of $\sim$20\,mK in a dilution refrigerator equipped with a 6/2/2\,T vector magnet. Two different measurement techniques are used in this work. (1) Conductance measurements are performed with standard dc technique. (2) For the rf measurements, resonators typically have resonance frequencies in the range of 250-450\,MHz. For resonators, the inductors have inductances of 300-730\,nH and a parasitic capacitance of $\sim$0.5\,pF. The acquisition time for each data point is typically 1\,ms. In each tunneling spectroscopy line trace, there are typically 201 or 401 data points. The rf signals are generated and demodulated by UHFLI from Zurich Instruments. Within the measurements presented in this work, the junction conductances on the two sides are kept at relatively low values (below 0.3\,$G_{0}$) in order to minimize voltage divider effects from serial resistances in the setup while assuring sensitivity to rf detection. An arbitrary waveform generator Tektronix 5014C generates the waveforms which serve as bias voltages.\\ 

\noindent\textbf{Correspondence between rf reflection and differential conductance.} In our data, zero-field rf reflection results can be converted into differential conductance. Converting rf reflection data at finite magnetic field is difficult due to inadequate calibration on resonators in field, which has little influence on this article as we focus on the energy of sub-gap states rather than conductance amplitude of these states. 

\noindent\textbf{Data analysis.} All details in data analysis are included in a jupyter notebook file (link is specified in section \textbf{Data availability}). \\       

\noindent\textbf{Data availability.} The authors declare that all relevant raw data together with analysis files are available at \url{https://doi.org/10.5281/zenodo.5938281}\\     


\section{Additional data}

\begin{figure*}[!t]
\centering
\includegraphics[width=0.65\linewidth]{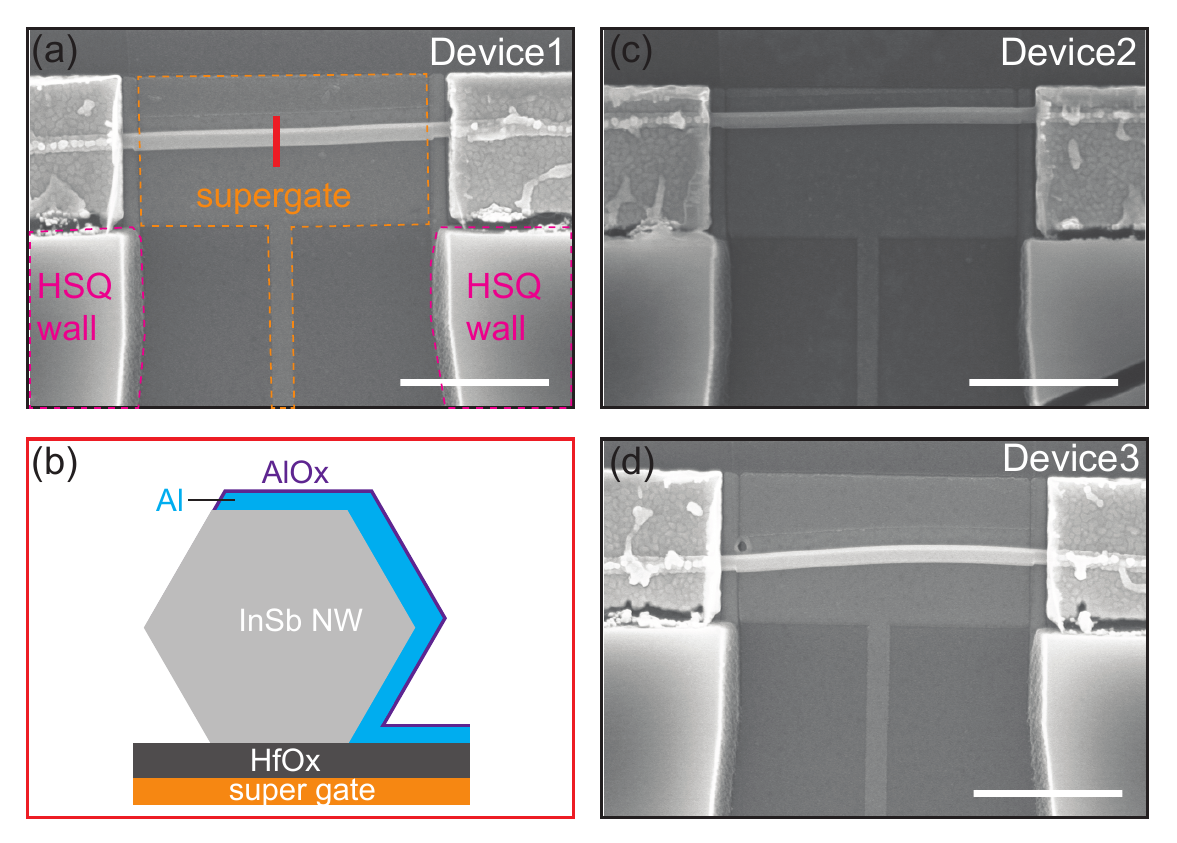}
\caption{Scanning-electron microscope (SEM) images of the three InSb-Al nanowire devices presented in this work (\textbf{(a)}, \textbf{(c)} and \textbf{(d)}). The length of the InSb-Al hybrids is $\sim$2\,$\mu$m. Scale bars in the pictures are 1\,$\mu$m. In \textbf{(a)}, the super gate and insulating nanostructures (HSQ wall) for the shadow-wall evaporation are outlined in orange and magenta, respectively. \textbf{(b)} Schematic cross-section of an InSb-Al nanowire hybrid.  
}\label{fig:S1}
\end{figure*}

\begin{figure*}[!t]
\centering
\includegraphics[width=0.9\linewidth]{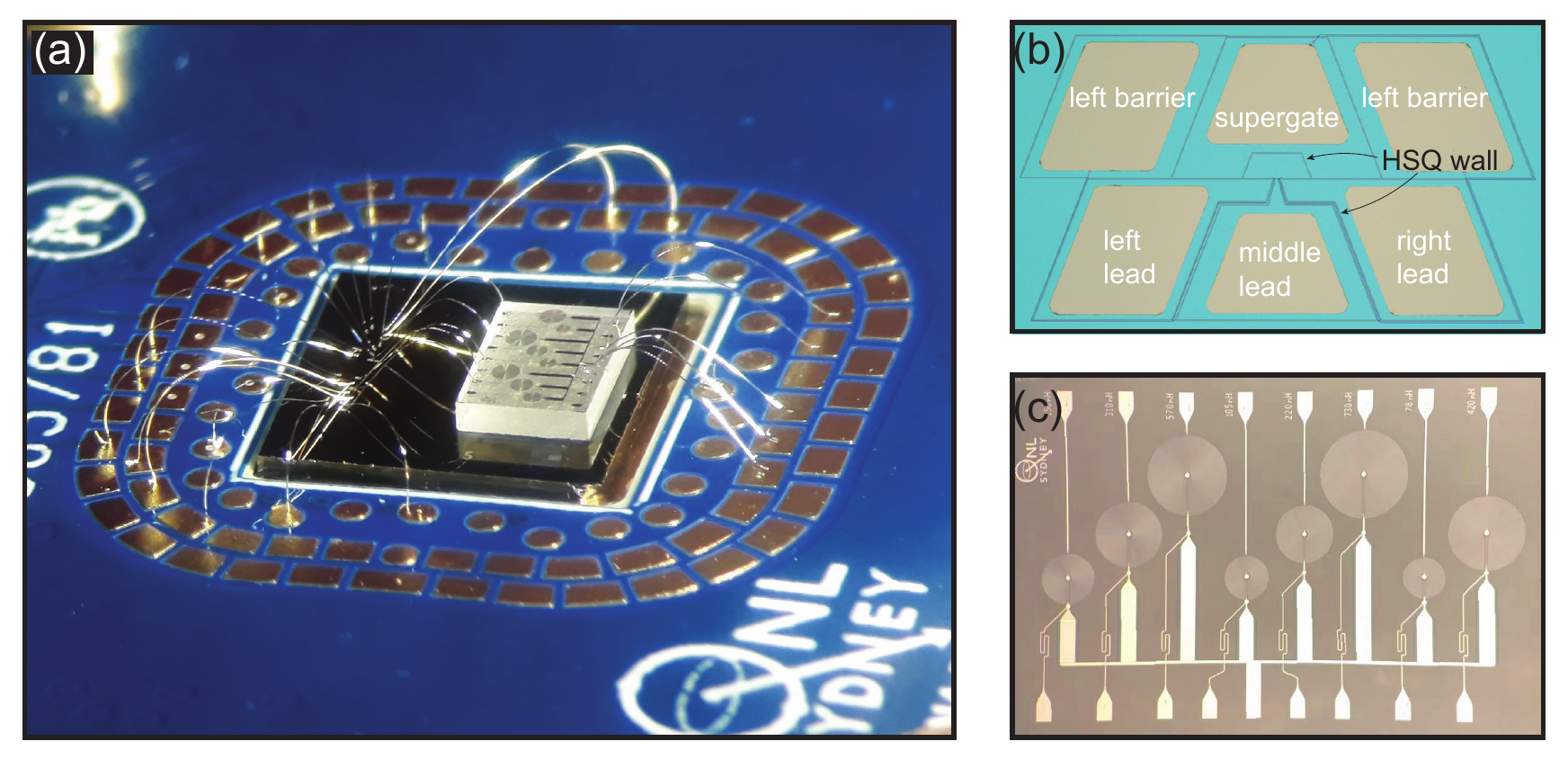}
\caption{\textbf{(a)} Optical image of a PCB board with a sample chip and a frequency multiplexing chip \cite{bib:Hornibrook2014} after bonding. \textbf{(b)} Optical image of a representative device. Bonding pads for different gates and leads are labelled. \textbf{(c)} Optical image of a frequency multiplexing chip.     
}\label{fig:S2}
\end{figure*}

\begin{figure*}[!t]
\centering
\includegraphics[width=0.9\linewidth]{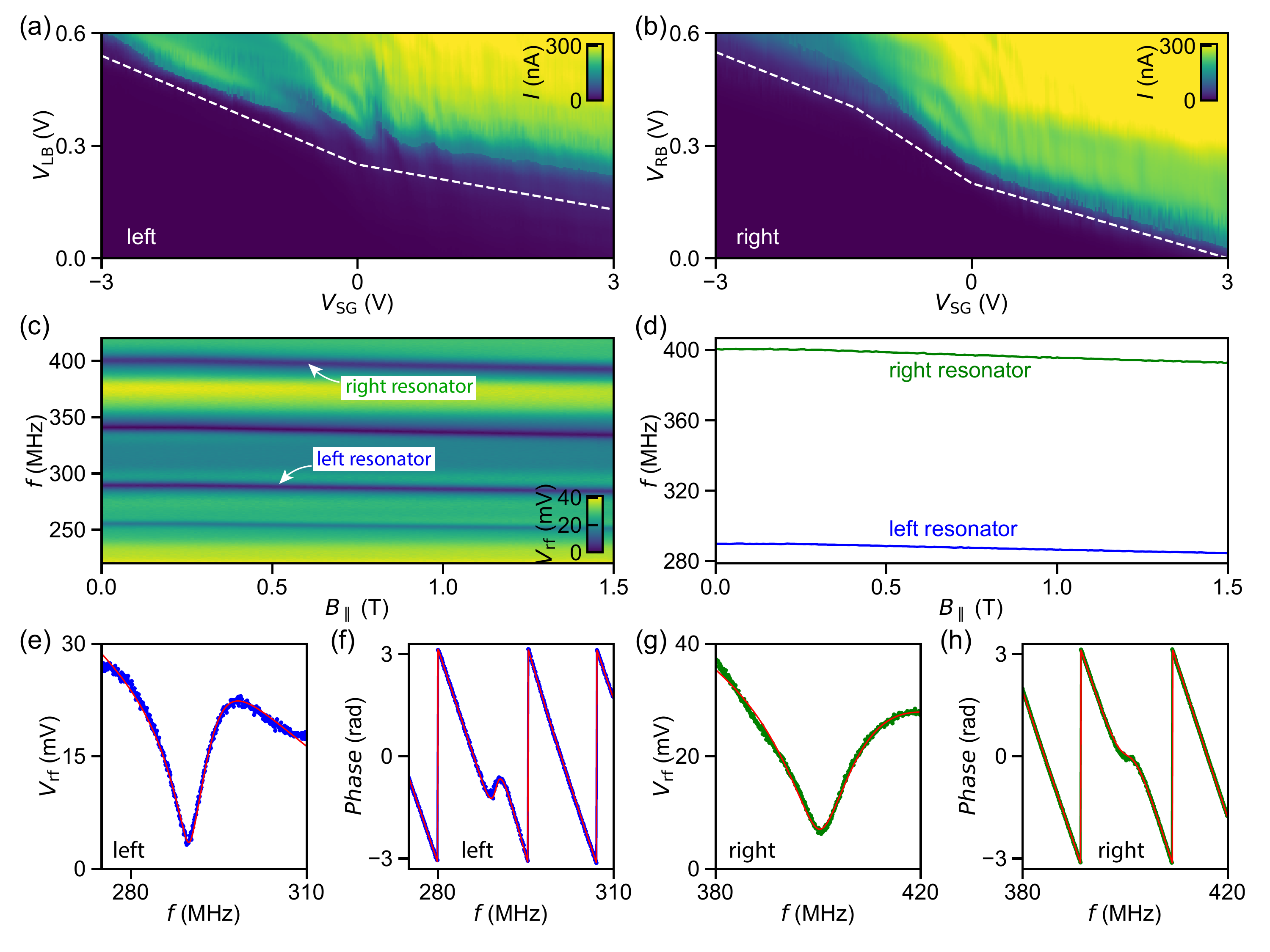}
\caption{Basic characteristics of \textbf{Device 1}. \textbf{(a)} Current through the left junction, $I$, as a function of barrier gate voltage $V_{\mathrm{LB}}$ and $V_{\mathrm{SG}}$ at 10\,mV bias voltage. \textbf{(b)} Current through the right junction, $I$, as a function of barrier gate voltage $V_{\mathrm{RB}}$ and $V_{\mathrm{SG}}$ at 10\,mV bias voltage. In Fig.3 in the main text, the super gate is swept while compensating the two barrier gate voltages by following the white dashed lines in \textbf{(a)} and \textbf{(b)}. \textbf{(c)} Dependence of the resonators on parallel magnetic field. Four resonators are visible in this panel, of which two are bonded to \textbf{Device 1}. \textbf{(d)} Extracted resonant frequencies of the left and right resonators as a function of $B_{\mathrm{||}}$. \textbf{(e)-(h)} Resonator reflection spectra (blue and green solid dots) and corresponding fitted curves (red solid lines) for two resonators based on a hanger superconducting resonator model \cite{bib:Khalil2012}. \textbf{(e)} and \textbf{(f)} show respectively the amplitude and phase of the left resonator. \textbf{(g)} and \textbf{(h)} show respectively the amplitude and phase of the right resonator. Internal quality factors of the two resonators, $Q_{\mathrm{L}}$ and $Q_{\mathrm{R}}$, are fitted to be $\sim$\,47 and $\sim$\,28, respectively.            
}\label{fig:S3}
\end{figure*}

\begin{figure*}[!t]
\centering
\includegraphics[width=0.9\linewidth]{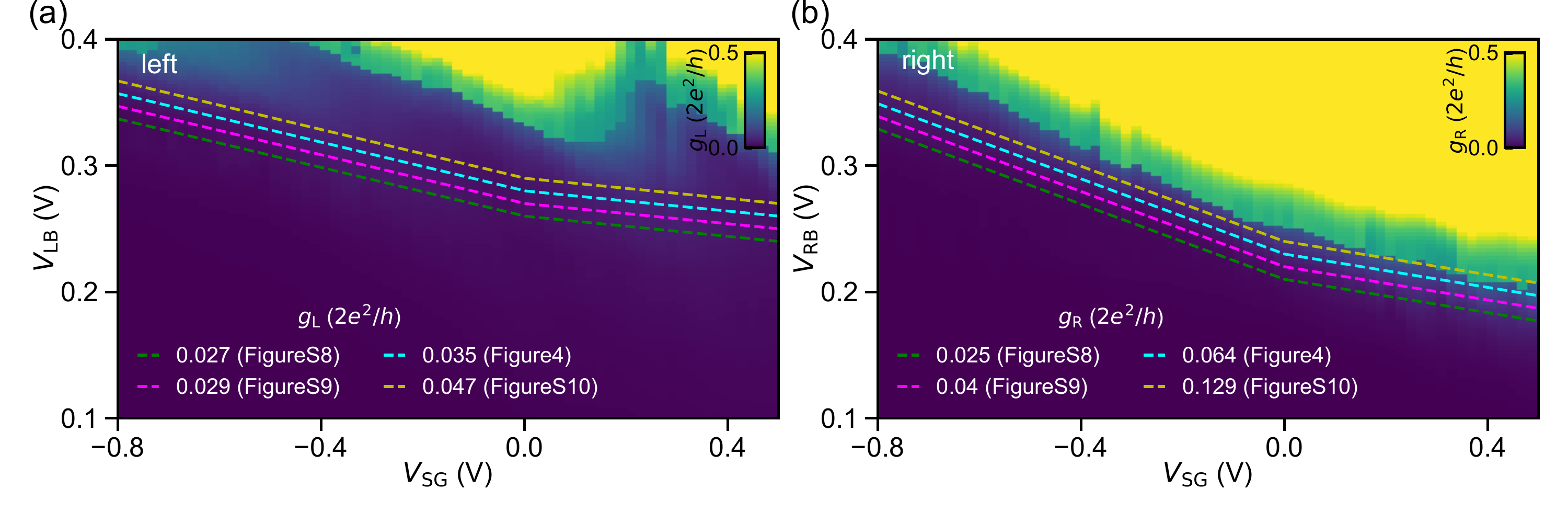}
\caption{Conductance measurements of the two junctions in \textbf{Device 1}, together with gate-compensation lines used in tunneling spectroscopy measurements. \textbf{(a)} Calculated conductance of the left junction, $g_{\mathrm{L}}$, from Fig.~\ref{fig:S3}(a) \textbf{(b)} Calculated conductance of the right junction, $g_{\mathrm{R}}$, from Fig.~\ref{fig:S3}(b). In \textbf{(a)} and \textbf{(b)}, dashed lines with different colors illustrate how gates are swept during corresponding measurements.     
}\label{fig:S4}
\end{figure*}

\begin{figure*}[!t]
\centering
\includegraphics[width=0.90\linewidth]{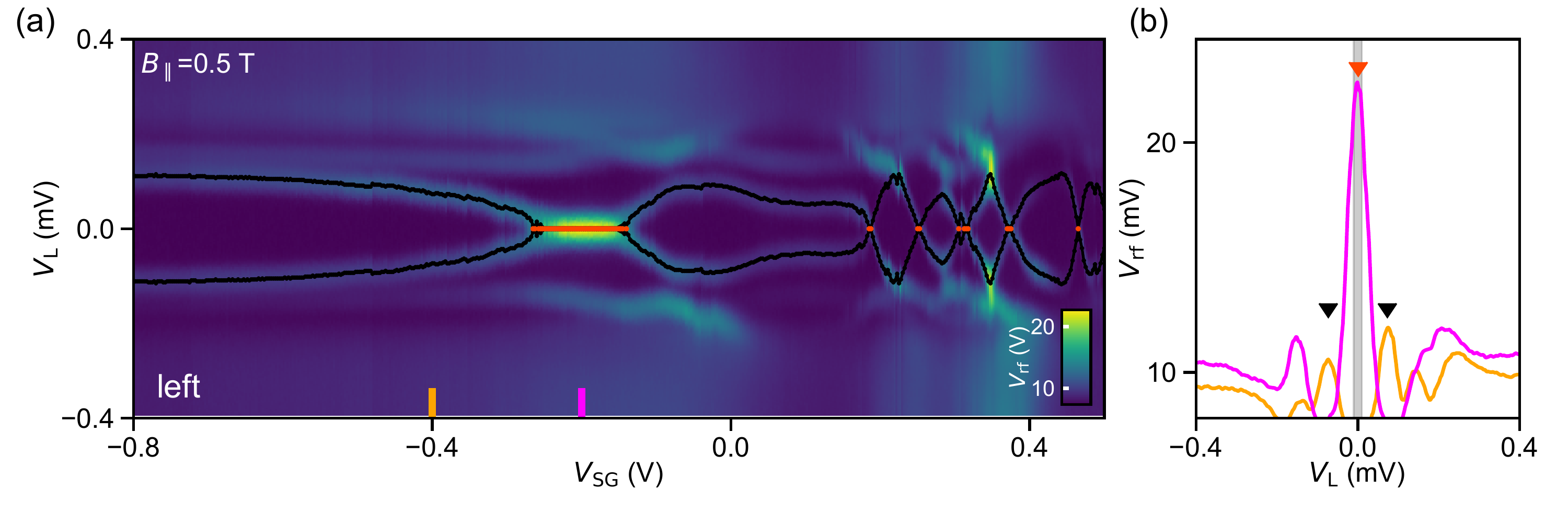}
\caption{Example of the energy extraction for LESs and ZESs. Data from \textbf{Device 1}. \textbf{(a)} Tunneling spectroscopy versus super gate voltage at $B_{\mathrm{||}}$ = 0.5\,T, taken from the data set of Fig.4 in the main text. Black and red dots mark the extracted LESs and ZESs, respectively. Line cuts, taken at $V_{\mathrm{SG}}$ = $\num{-0.4}$\,V (orange) and $V_{\mathrm{SG}}$ = $\num{-0.2}$\,V (magenta), are shown in \textbf{(b)}. Black triangles mark the position of the extracted LESs, whilst red triangle marks the position of the extracted ZESs. The grey area displays the bias window from $\num{-10}$\,$\mu$V to $\num{10}$\,$\mu$V used for the extraction of ZESs.   
}\label{fig:S5}
\end{figure*}

\begin{figure*}[!t]
\centering
\includegraphics[width=0.5\linewidth]{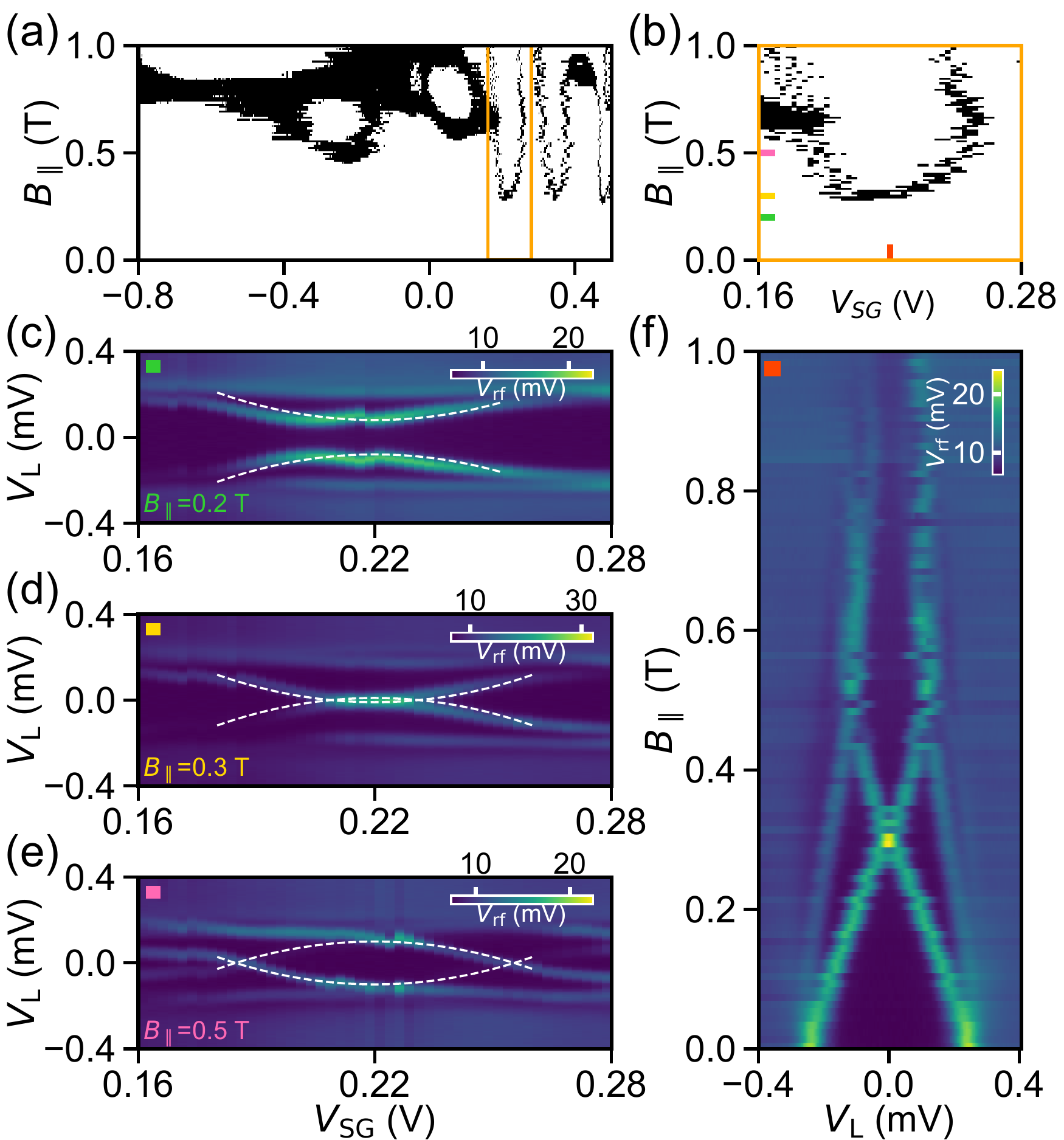}
\caption{Another example of parabolic patterns in ZES diagrams for \textbf{Device 1}. \textbf{(a)} ZES diagram for the left junction (same as Fig.4(g) in the main text). The region marked by the orange rectangle is shown in \textbf{(b)}. Three line cuts are made at different magnetic fields and corresponding tunneling spectroscopy data is shown in \textbf{(c)-(e)}. White dashed lines serve as a guide to the eye for a pair of sub-gap levels. \textbf{(f)} Evolution of sub-gap levels in magnetic field at $V_{\mathrm{SG}}$ = 0.22\,V. Together with Fig. 6 in the main text, we show that parabolic patterns in ZES diagrams, consistent with onset of a topological phase in short hybrid nanowires \cite{bib:Sarma2012}, can be fully explained by Zeeman-driven Andreev level splitting in a quantum dot$-$superconducting lead architecture \cite{bib:Lee2014}.
}\label{fig:S6}
\end{figure*}

\begin{figure*}[!t]
\centering
\includegraphics[width=1.0\linewidth]{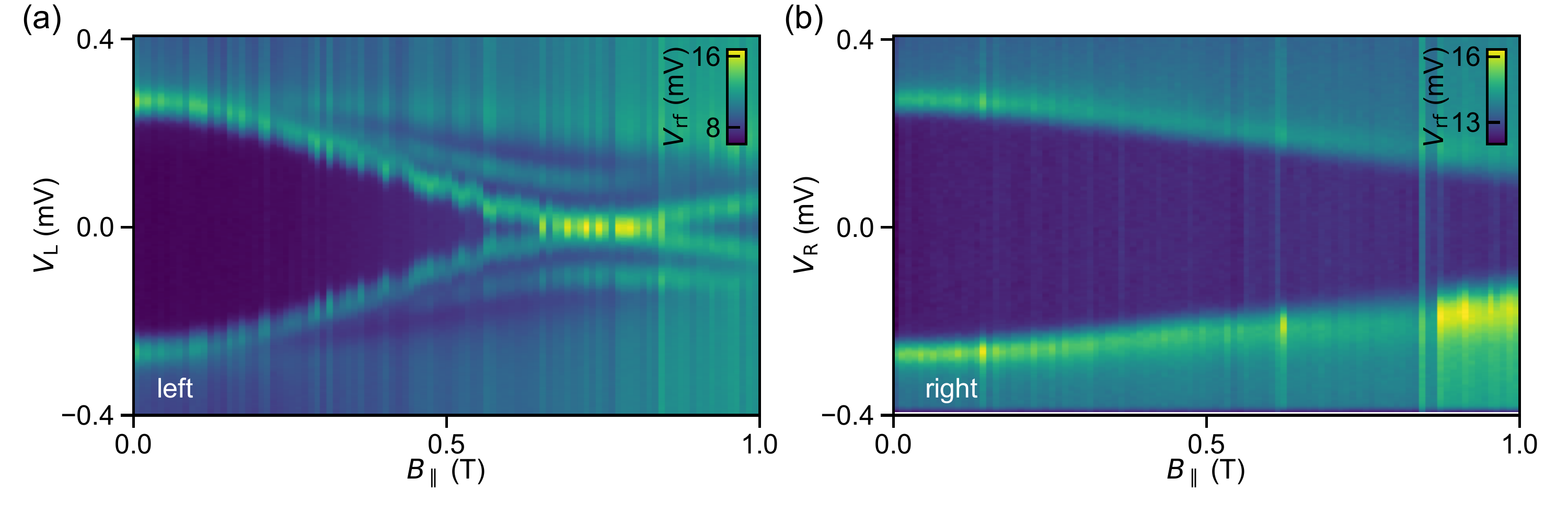}
\caption{Tunneling spectroscopy results at $V_{\mathrm{SG}}$ = $\num{-0.5}$\,V for the left \textbf{(a)} and right \textbf{(b)} junction. For the left side, sub-gap states are present whilst they are absent for the right side. Data is taken from the data set of Fig.4 in the main text (\textbf{Device 1}).   
}\label{fig:S7}
\end{figure*}

\begin{figure*}[!t]
\centering
\includegraphics[width=0.90\linewidth]{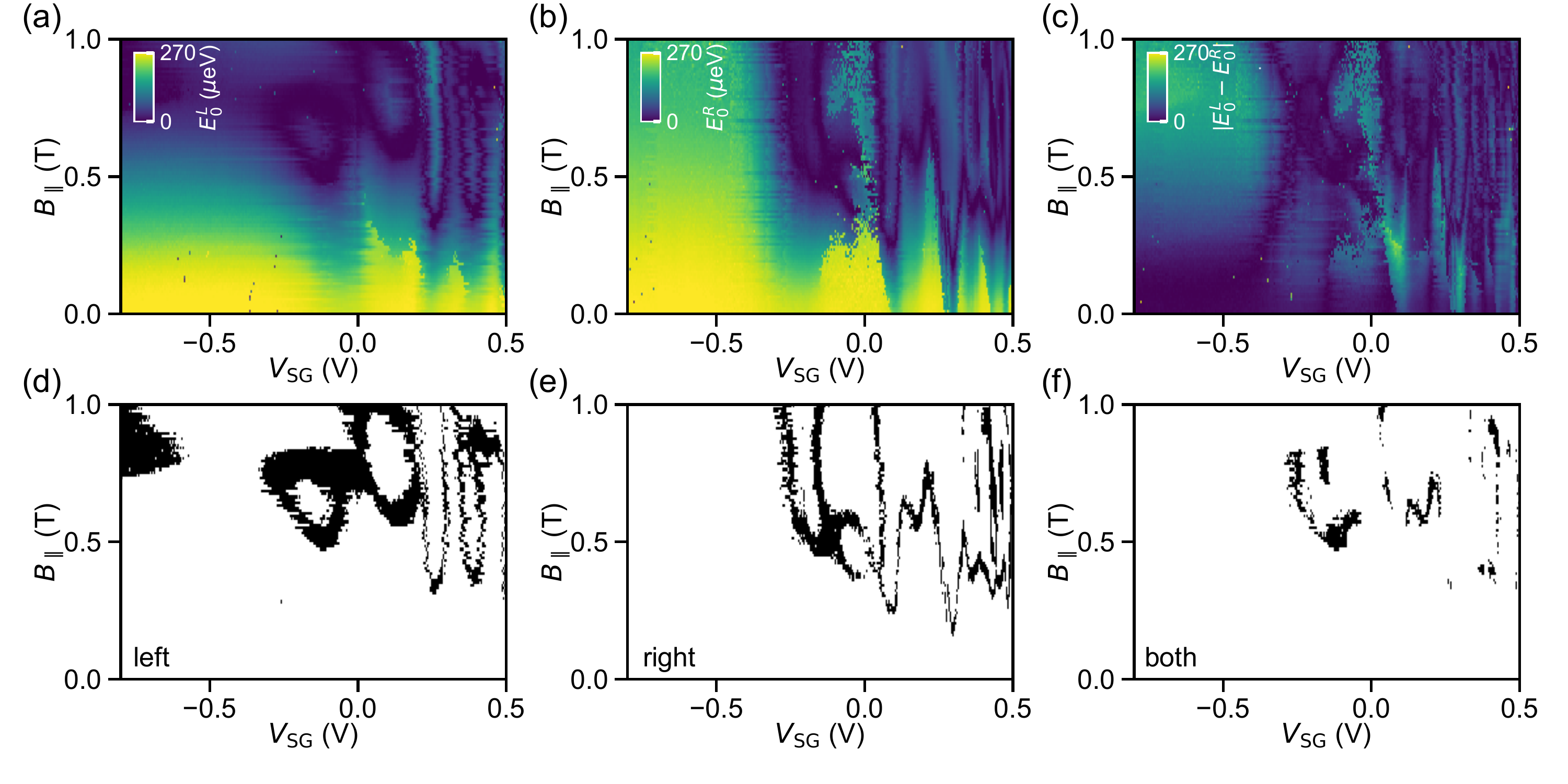}
\caption{LES and ZES diagrams for $g_{\mathrm{L}}\sim$\,0.027\,$G_{\mathrm{0}}$, $g_{\mathrm{R}}\sim$\,0.025\,$G_{\mathrm{0}}$. Data comes from \textbf{Device 1}. Gates are swept by following the green dashed lines in Fig.~\ref{fig:S4}. Compared with Fig. 4 in the main text, both barrier gates are shifted to more negative value by 20 mV, in order to confirm the stability of the LESs and ZESs against varying tunneling transparency. \textbf{(a)} LES diagram in $V_{\mathrm{SG}}$$-$$B_{\mathrm{||}}$ space on the left end of the hybrid nanowire. \textbf{(b)} LES diagram in $V_{\mathrm{SG}}$$-$$B_{\mathrm{||}}$ space on the right end of the hybrid nanowire. \textbf{(c)} Energy difference between LESs probed on the two sides. \textbf{(d)} and \textbf{(e)} ZES diagrams for the left and right junctions, respectively. \textbf{(f)} Diagram of coexisting ZESs on both ends. In \textbf{(a)-(f)}, each panel includes 522$\times$101 pixels. The data set is taken in $\sim$35\,hours.
}\label{fig:S8}
\end{figure*}

\begin{figure*}[!t]
\centering
\includegraphics[width=0.90\linewidth]{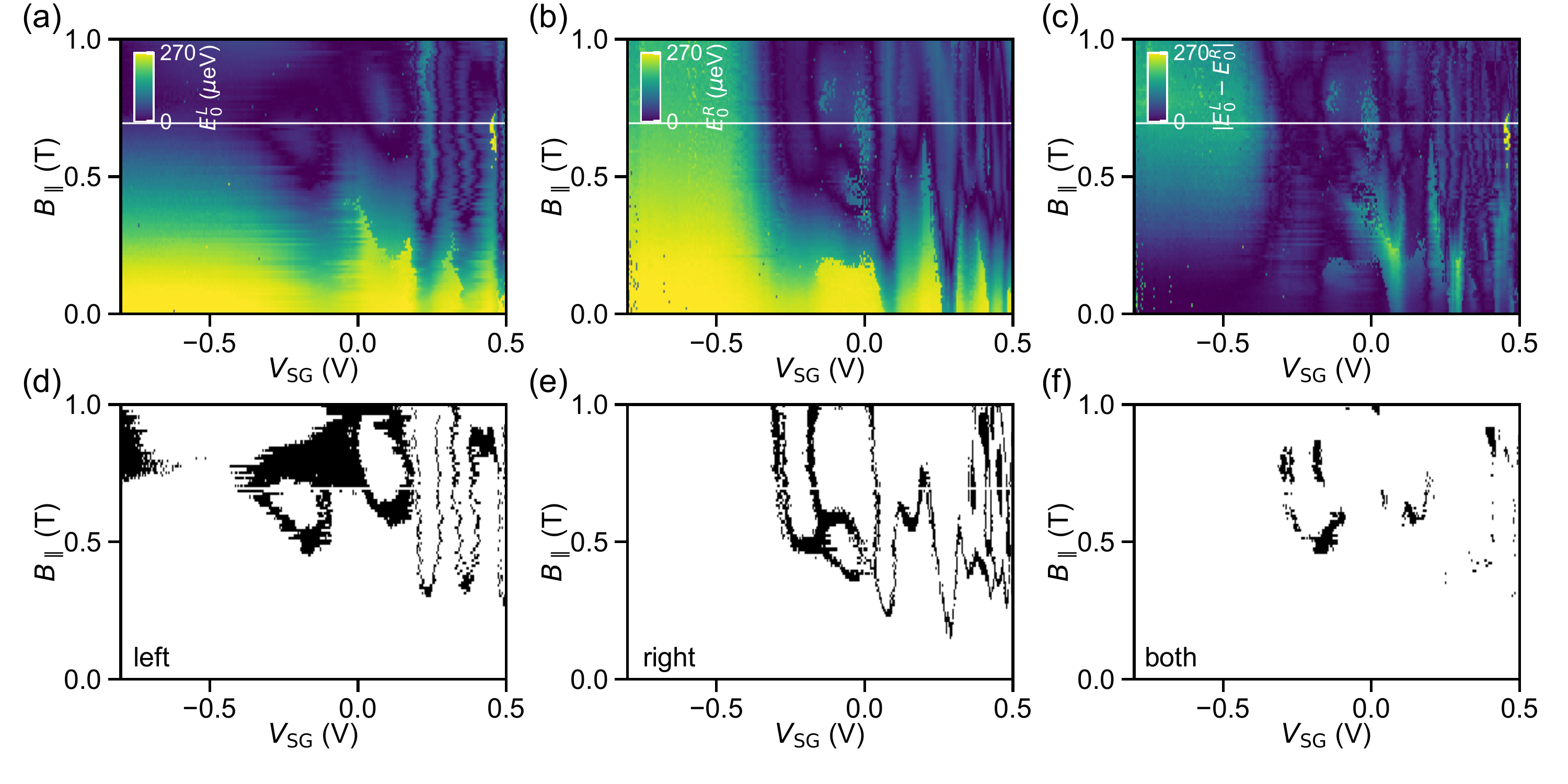}
\caption{LES and ZES diagrams for $g_{\mathrm{L}}\sim$\,0.029\,$G_{\mathrm{0}}$, $g_{\mathrm{R}}\sim$\,0.04\,$G_{\mathrm{0}}$. Data comes from \textbf{Device 1}. Gates are swept by following the magenta dashed lines in Fig.~\ref{fig:S4}. Compared with Fig. 4 in the main text, both barrier gates are shifted to more negative value by 10 mV, in order to confirm the stability of the LESs and ZESs against varying tunneling transparency. \textbf{(a)} LES diagram in super gate-magnetic field space on the left end of the hybrid nanowire. \textbf{(b)} LES diagram in super gate-magnetic field space on the right end of the hybrid nanowire. \textbf{(c)} Energy difference between ZESs probed on the two sides. \textbf{(d)} and \textbf{(e)} ZES diagrams for the left and right junctions, respectively. \textbf{(f)} Diagram of coexistent ZESs on both ends. In \textbf{(a)-(f)}, each panel includes 522$\times$100 pixels. The dataset is taken in $\sim$34.5 hours. Note that compared with data sets in Fig.4 in the main text, Fig.~\ref{fig:S8} and Fig.~\ref{fig:S10}, in this data set the results at $B_{\mathrm{||}}$ = 0.69\,T was not taken because of technical errors. In \textbf{(a)-(c)}, data at $B_{\mathrm{||}}$ = 0.69\,T is set to NaN and panels display a white line at this field value.  
}\label{fig:S9}
\end{figure*}

\begin{figure*}[!t]
\centering
\includegraphics[width=0.90\linewidth]{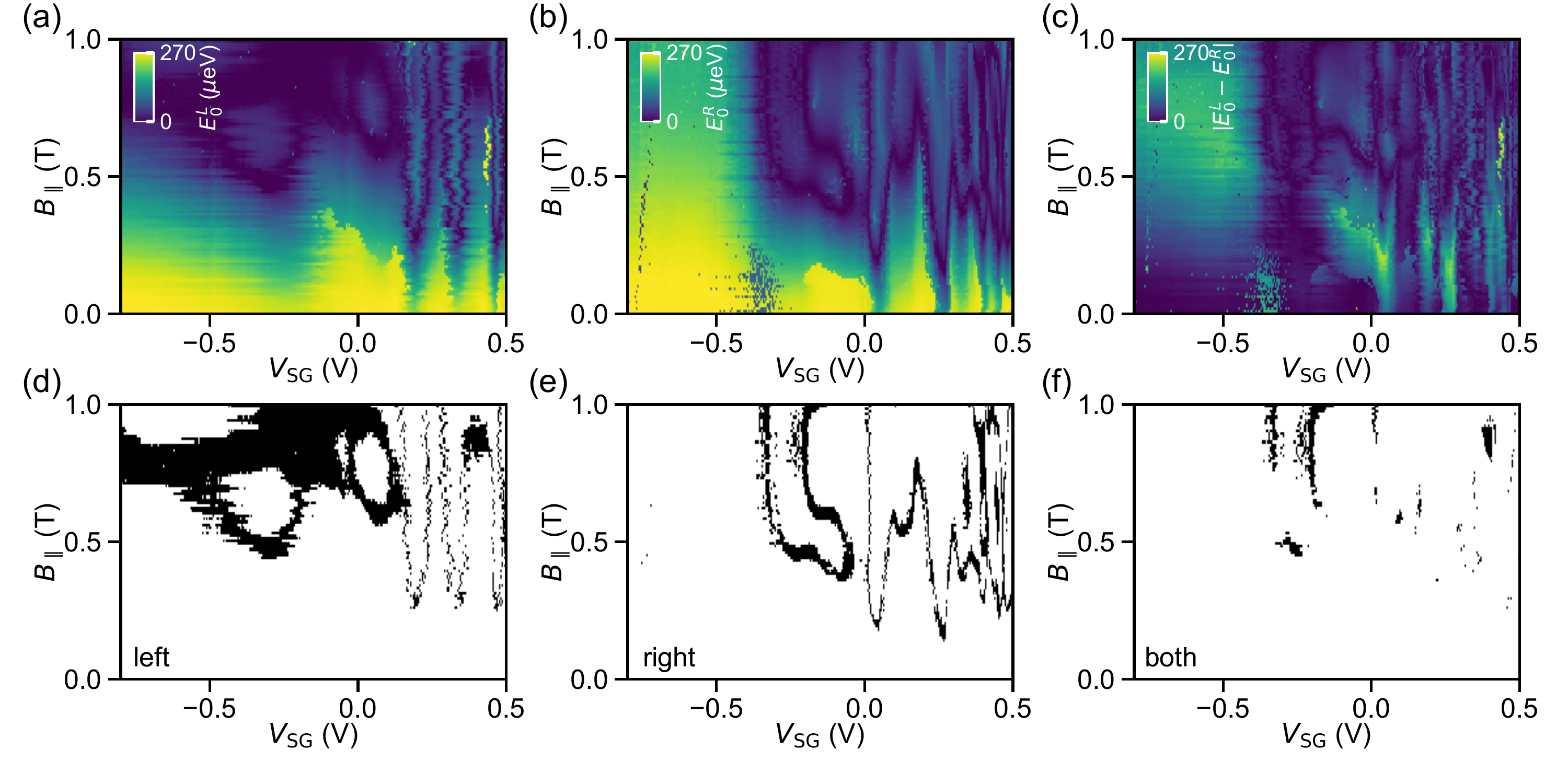}
\caption{LES and ZES diagrams for $g_{\mathrm{R}}\sim$\,0.047\,$G_{\mathrm{0}}$, $g_{\mathrm{R}}\sim$\,0.129\,$G_{\mathrm{0}}$. Data comes from \textbf{Device 1}. Gates are swept by following the yellow dashed lines in Fig.~\ref{fig:S4}. Compared with Fig. 4 in the main text, both barrier gates are shifted to more positive value by 10 mV, in order to confirm the stability of the LESs and ZESs against varying tunneling transparency. \textbf{(a)} LES diagram in super gate-magnetic field space on the left end of the hybrid nanowire. \textbf{(b)} LES diagram in super gate-magnetic field space on the right end of the hybrid nanowire. \textbf{(c)} Energy difference between ZESs probed on both ends. \textbf{(d)} and \textbf{(e)} ZES diagrams for left and right junctions, respectively. \textbf{(f)} Diagram of coexistent ZESs on both ends. In \textbf{(a)-(f)}, each panel includes 522$\times$101 pixels. The data set is taken in $\sim$35\,hours.
}\label{fig:S10}
\end{figure*}

\begin{figure*}[!t]
\centering
\includegraphics[width=0.90\linewidth]{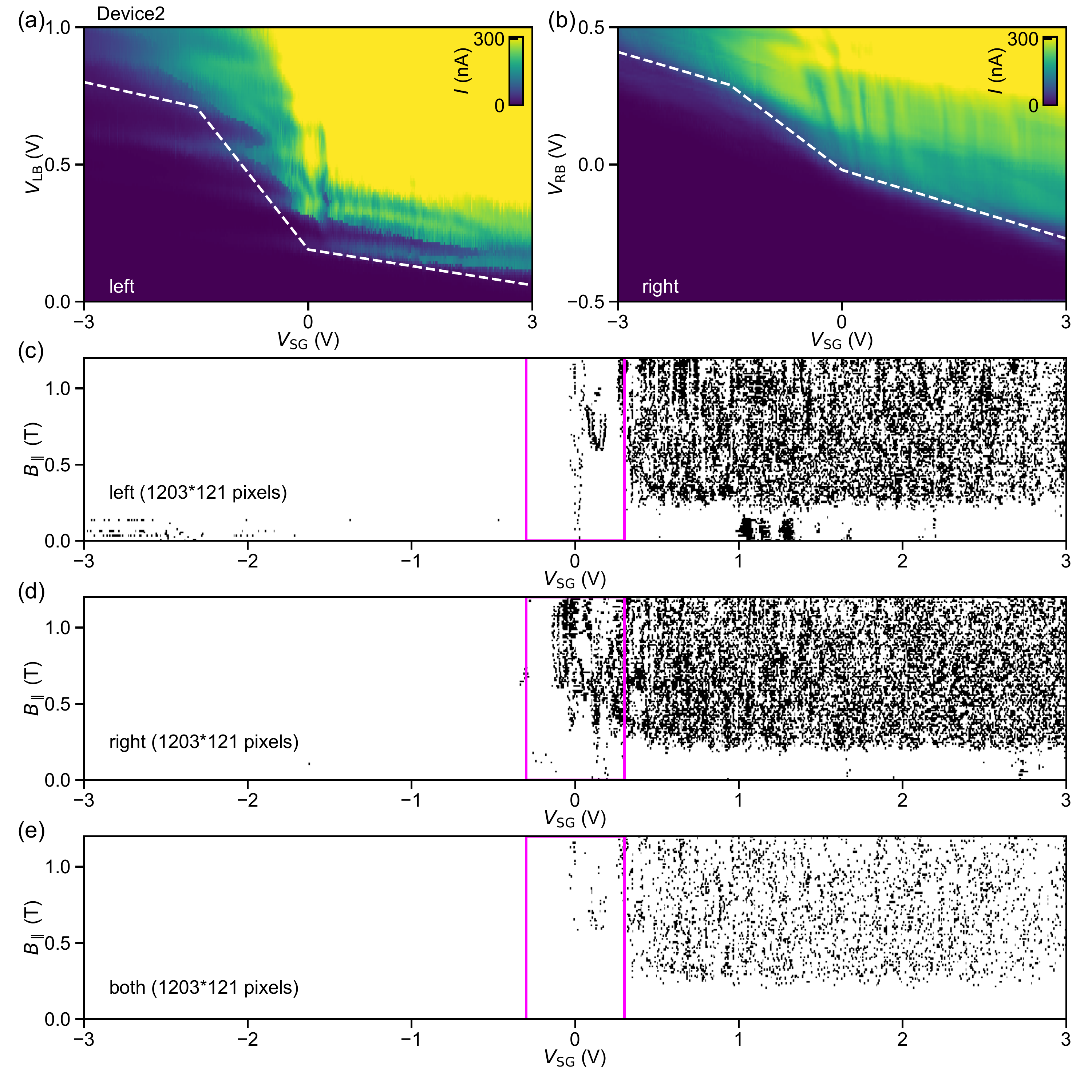}
\caption{Cross-talk properties of \textbf{Device 2} and large-scale ZES diagrams. \textbf{(a)} Current through the left junction, $I$, as a function of barrier gate voltage  $V_{\mathrm{LB}}$ and $V_{\mathrm{SG}}$ at 10\,mV bias voltage. \textbf{(b)} Current through the right junction, $I$, as a function of barrier gate voltage $V_{\mathrm{RB}}$ and $V_{\mathrm{SG}}$ at 10\,mV bias voltage. ZES diagrams in parameter space of super gate voltage $V_{\mathrm{SG}}$ and parallel magnetic field $B_{\mathrm{||}}$ for the left junction \textbf{(c)}, right junction \textbf{(d)}, and coexistence in both junctions \textbf{(e)}. Gates are scanned by following the white dashed lines in \textbf{(a)} and \textbf{(b)}. In \textbf{(c)-(e)}, each panel includes 1203$\times$121 pixels. The data set is taken in $\sim$\,96.8\,hours. Similar as Fig. 3 in the main text (\textbf{Device 1}), three distinct regimes along $V_{\mathrm{SG}}$ can be identified and the intermediate supergate regime (marked by magenta rectangles) is further investigated in Fig.~\ref{fig:S12}. 
}\label{fig:S11}
\end{figure*}

\begin{figure*}[!t]
\centering
\includegraphics[width=0.80\linewidth]{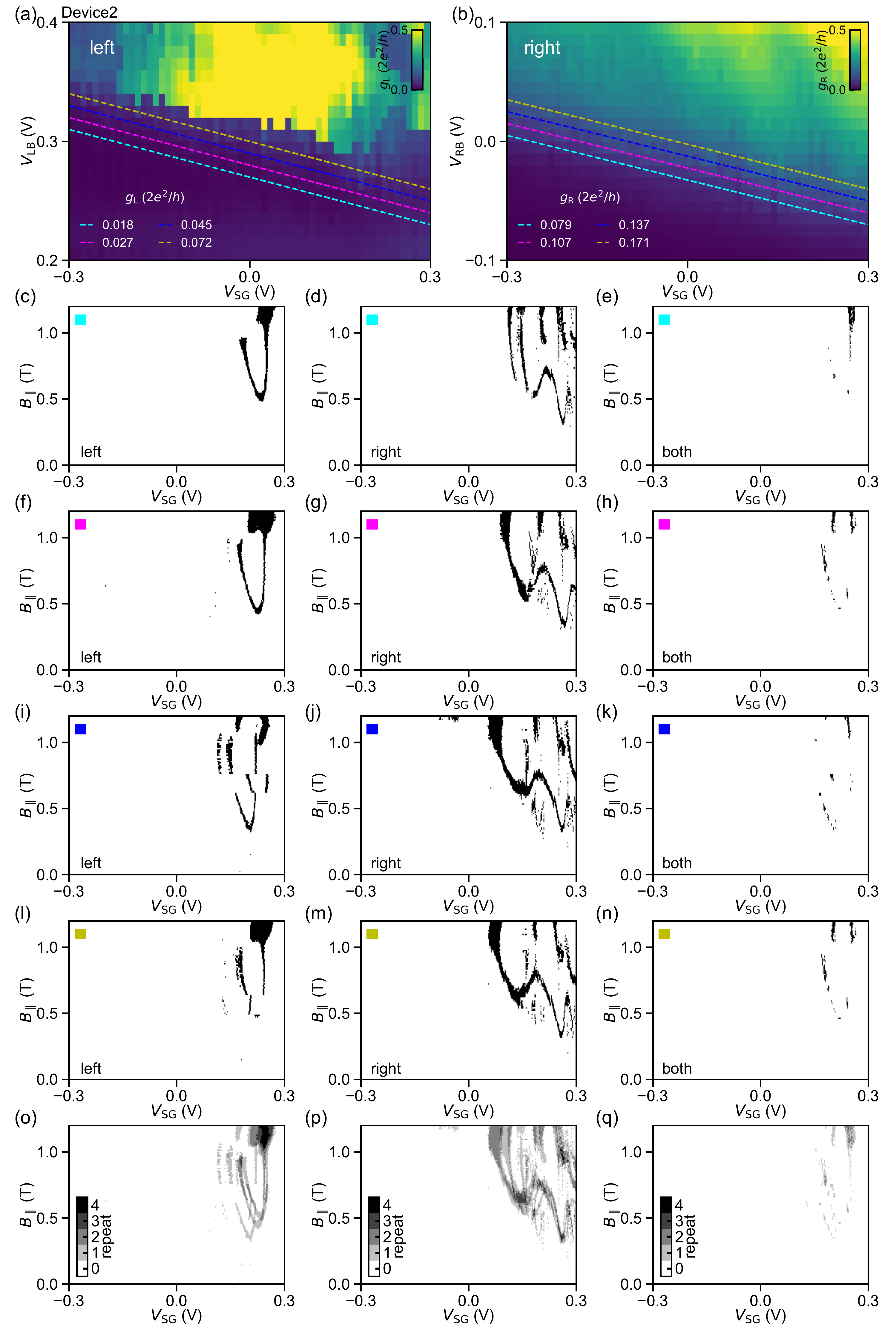}
\caption{Conductance measurements of the two junctions in \textbf{Device 2}, together with gate-compensation lines used in tunneling spectroscopy measurements. \textbf{(a)} Calculated conductance of the left junction, $g_{\mathrm{L}}$, from Fig.~\ref{fig:S11}(a) \textbf{(b)} Calculated conductance of the right junction, $g_{\mathrm{R}}$, from Fig.~\ref{fig:S11}(b). In \textbf{(a)} and \textbf{(b)}, dashed lines with different colors illustrate how gates are swept during corresponding measurements. The dashed lines have a difference of 10 mV in each barrier gate with respect to each other. \textbf{(c)}-\textbf{(n)} ZES diagrams at various barrier gate settings. \textbf{(c)}, \textbf{(f)}, \textbf{(i)}, and \textbf{(l)} show results for the left junction. \textbf{(d)}, \textbf{(g)}, \textbf{(j)}, and \textbf{(m)} show results for right junction. \textbf{(e)}, \textbf{(h)}, \textbf{(k)}, and \textbf{(n)} show the diagram of coexistent ZESs on both sides. \textbf{(c)}-\textbf{(e)} are taken at $g_{\mathrm{L}}\sim$\,0.018\,$G_{\mathrm{0}}$ and $g_{\mathrm{R}}\sim$\,0.079\,$G_{\mathrm{0}}$, marked with a cyan color. \textbf{(f)}-\textbf{(h)} are taken at $g_{\mathrm{L}}\sim$\,0.027\,$G_{\mathrm{0}}$ and $g_{\mathrm{R}}\sim$\,0.107\,$G_{\mathrm{0}}$, marked with a magenta color. \textbf{(i)}-\textbf{(k)} are taken at $g_{\mathrm{L}}\sim$\,0.045\,$G_{\mathrm{0}}$ and $g_{\mathrm{R}}\sim$\,0.137\,$G_{\mathrm{0}}$, marked with a blue color. \textbf{(l)}-\textbf{(n)} are taken at $g_{\mathrm{L}}\sim$\,0.072\,$G_{\mathrm{0}}$ and $g_{\mathrm{R}}\sim$\,0.171\,$G_{\mathrm{0}}$, marked with a yellow color. From these data sets, overlapping ZES diagrams are constructed which are shown for the left \textbf{(o)} and right \textbf{(p)} junction, as well as coexistence in both junctions \textbf{(q)}. In \textbf{(c)-(n)}, each panel includes 601$\times$121 pixels. The total data set is taken in $\sim$\,193.6\,hours.}\label{fig:S12}
\end{figure*}

\begin{figure*}[!t]
\centering
\includegraphics[width=0.90\linewidth]{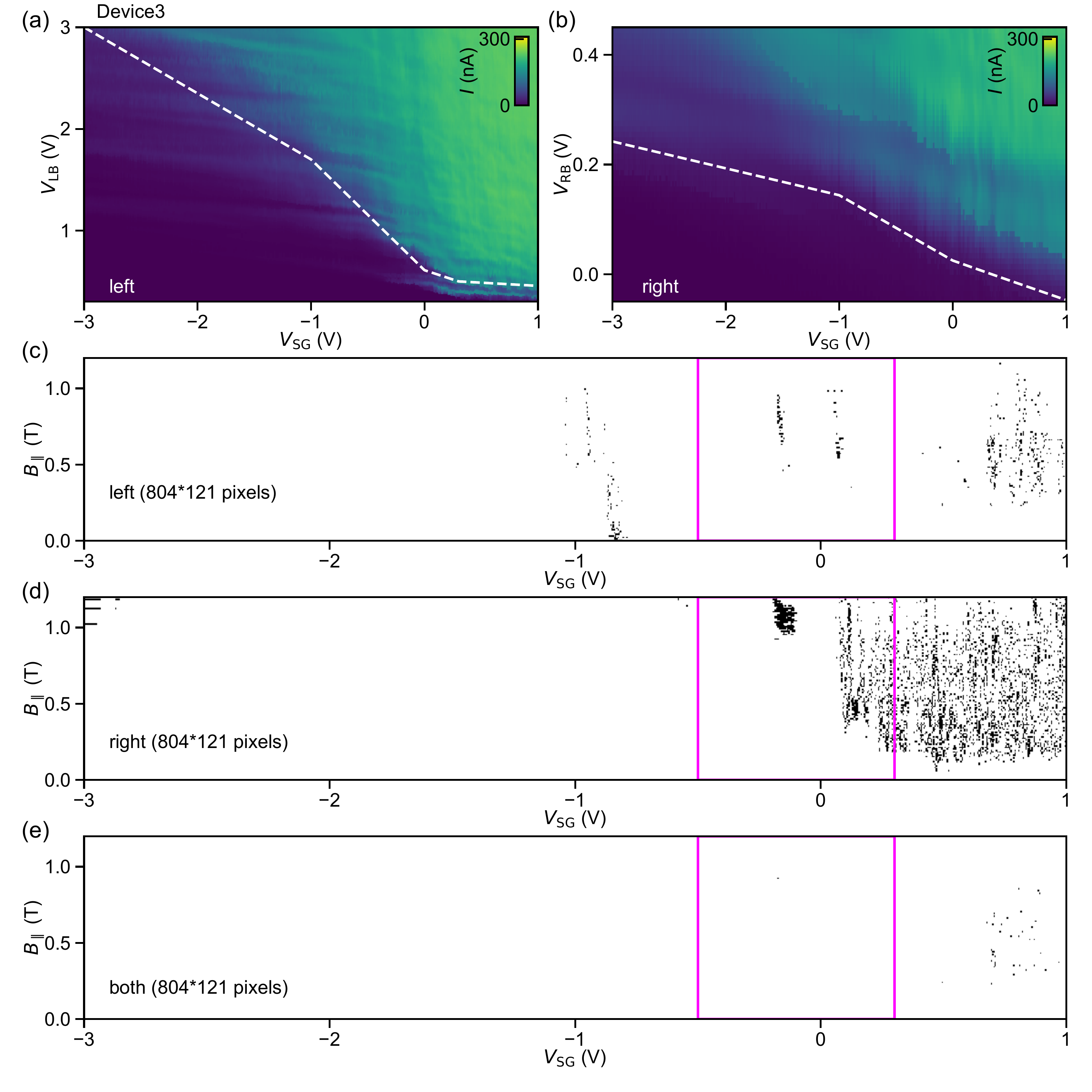}
\caption{Cross-talk properties of \textbf{Device 3} and large-scale ZES diagrams. \textbf{(a)} Current through the left junction, $I$, as a function of barrier gate voltage  $V_{\mathrm{LB}}$ and $V_{\mathrm{SG}}$ at 5\,mV bias voltage. \textbf{(b)} Current through the right junction, $I$, as a function of barrier gate voltage $V_{\mathrm{RB}}$ and $V_{\mathrm{SG}}$ at 5\,mV bias voltage. ZES diagrams in parameter space of super gate voltage $V_{\mathrm{SG}}$ and parallel magnetic field $B_{\mathrm{||}}$ for the left junction \textbf{(c)}, right junction \textbf{(d)}, and their intersection \textbf{(e)}. Gates are scanned by following the white dashed lines in \textbf{(a)} and \textbf{(b)}. In \textbf{(c)-(e)}, each panel includes 804$\times$121 pixels. The data set is taken in $\sim$\,42.3\,hours. Similar as Fig. 3 in the main text (\textbf{Device 1}) and Fig.~\ref{fig:S11} (textbf{Device 2}), there are three different regimes along $V_{\mathrm{SG}}$ for right side, while left side does not have such clear characteristics. Nonethless, we still focus on similar intermediate super gate region (within magenta rectangles) and do further investigation in Fig.~\ref{fig:S14}.
}\label{fig:S13}
\end{figure*}

\begin{figure*}[!t]
\centering
\includegraphics[width=0.90\linewidth]{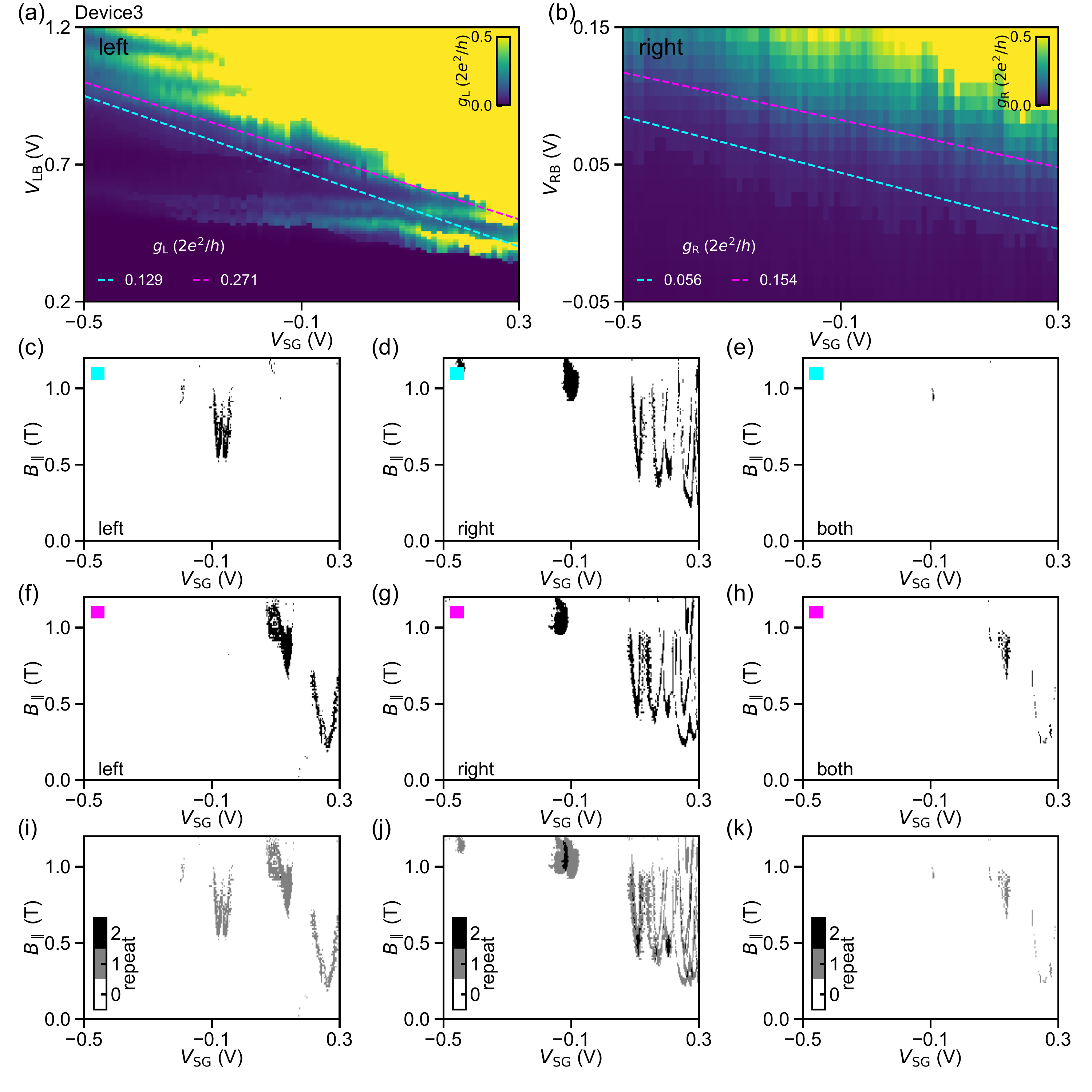}
\caption{Conductance measurements of the two junctions in \textbf{Device 3}, together with gate-compensation lines used in tunneling spectroscopy measurements. \textbf{(a)} Calculated conductance of the left junction, g$_{L}$, from Fig.~\ref{fig:S13}(a) \textbf{(b)} Calculated conductance of the right junction, g$_{R}$, from Fig.~\ref{fig:S13}(b). In \textbf{(a)} and \textbf{(b)}, dashed lines with different colors illustrate how gates are swept during corresponding measurements. \textbf{(c)}-\textbf{(h)} ZES diagrams at various barrier gate settings. \textbf{(c)} and \textbf{(f)} show results for the left junction. \textbf{(d)} and \textbf{(g)} show results for right junction. \textbf{(e)} and \textbf{(h)} show the diagram of coexistent ZESs on both sides. \textbf{(c)}-\textbf{(e)} are taken at $g_{\mathrm{L}}\sim$\,0.129\,$G_{\mathrm{0}}$ and $g_{\mathrm{R}}\sim$\,0.056\,$G_{\mathrm{0}}$, marked with a cyan color. \textbf{(f)}-\textbf{(h)} are taken at $g_{\mathrm{L}}\sim$\,0.271\,$G_{\mathrm{0}}$ and $g_{\mathrm{R}}\sim$\,0.154\,$G_{\mathrm{0}}$, marked with a magenta color. From these data sets, overlapping ZES diagrams are constructed which are shown for the left \textbf{(i)} and right \textbf{(j)} junction, as well as coexistence in both junctions \textbf{(k)}. In \textbf{(c)-(k)}, each panel includes 601$\times$121 pixels. The total data set is taken in $\sim$\,83\,hours. Notably, left side has dramatic changes in ZES diagrams with varying barrier gates (see \textbf{(c)} and \textbf{(f)}), while the right side just undergoes moderate changes (see \textbf{(d)} and \textbf{(g)}). The origins could be found from conductance measurement in \textbf{(a)} and \textbf{(b)}. Conductance of the left junction has quite distinct modulations along cyan and magenta dashed line, whereas for the right side conductance has similar modulations along two dashed lines.  
}\label{fig:S14}
\end{figure*}

